\documentclass[aps,amsmath,twocolumn,amssymb,floatfix,showpacs,superscriptaddress,nofootinbib,longbibliography]{revtex4-1}
\usepackage{braket}
\usepackage[dvipsnames]{xcolor}
\usepackage{float}
\usepackage{subfigure}
\usepackage[colorlinks=true,linktoc=page,linkcolor=magenta,citecolor=blue,urlcolor=purple]{hyperref}

\newcommand{\vect}[1]{\boldsymbol{\mathrm{#1}}}

\newcommand{\ie}{{\it i.e.,\,\,}}

\newcommand\bea{\begin{eqnarray}}
\newcommand\eea{\end{eqnarray}}
\newcommand\beq{\begin{equation}}  
\newcommand\eeq{\end{equation}}

\newcommand{\non}{\nonumber}  
\usepackage[normalem]{ulem}
\definecolor{lime}{HTML}{A6CE39}
\usepackage{sidecap,tikz}
\DeclareRobustCommand{\orcidicon}{\hspace{-1.0mm}
	\begin{tikzpicture}
		\draw[lime, fill=lime] (0.0,0.0) 
		circle [radius=0.15] 
		node[white] {{\fontfamily{qag}\selectfont \tiny \,ID}};
		\draw[white, fill=white] (-0.0525,0.095) 
		circle [radius=0.007];
	\end{tikzpicture}
	\hspace{-3.0mm}
}
\foreach \x in {A, ..., Z}{\expandafter\xdef\csname orcid\x\endcsname{\noexpand\href{https://orcid.org/\csname orcidauthor\x\endcsname}
		{\noexpand\orcidicon}}
}

\begin{document}


\title{Dynamical construction of quadrupolar and octupolar topological superconductors}

\author{Arnob Kumar Ghosh\orcidA{}}
\email{arnob@iopb.res.in}
\affiliation{Institute of Physics, Sachivalaya Marg, Bhubaneswar-751005, India}
\affiliation{Homi Bhabha National Institute, Training School Complex, Anushakti Nagar, Mumbai 400094, India}
\author{Tanay Nag\orcidB{}}
\email{tnag@physik.rwth-aachen.de}
\affiliation{Institut f\"ur Theorie der Statistischen Physik, RWTH Aachen University, 52056 Aachen, Germany}
\author{Arijit Saha\orcidC{}}
\email{arijit@iopb.res.in}
\affiliation{Institute of Physics, Sachivalaya Marg, Bhubaneswar-751005, India}
\affiliation{Homi Bhabha National Institute, Training School Complex, Anushakti Nagar, Mumbai 400094, India}

\begin{abstract}
	We propose a three-step periodic drive protocol to engineer two-dimensional~(2D) Floquet quadrupole superconductors and three-dimensional~(3D) Floquet octupole superconductors hosting zero-dimensional Majorana corner modes~(MCMs), based on unconventional $d$-wave superconductivity. Remarkably, the driven system conceives four phases with only $0$ MCMs, no MCMs, only anomalous $\pi$ MCMs, and both regular $0$ and anomalous $\pi$ MCMs. To circumvent the subtle issue of characterizing $0$ and $\pi$ MCMs separately, we employ the periodized evolution operator to architect the dynamical invariants, namely quadrupole and octupole motion in 2D and 3D, respectively, that can distinguish different higher order topological phases unambiguously. 
Our study paves the way for the realization of dynamical quadrupolar and octupolar topological superconductors. 
\end{abstract}

\maketitle

\section{Introduction}
Topological superconductors~(TSCs) hosting Majorana zero modes (MZMs) have been 
the corner stone for the last two decades due to their potential application in topological quantum computations utilising non-Abelian statistics~\cite{Kitaev_2001,Ivanov2001,nayak08,qi2011topological}. The quest for TSC emerges
following the elegant proposal by Kitaev~\cite{Kitaev_2001}, and the idea by Fu and Kane~\cite{Fu2008} that emphasized the realization of the MZMs on the two-dimensional~(2D) surface of a three-dimensional~(3D) topological insulator~(TI), in proximity to an $s$-wave superconductor and magnetic insulator. Very recently, the advent of generalized bulk boundary correspondence~(BBC) in the higher-order topological~(HOT) phase~\cite{benalcazar2017,benalcazarprb2017,Song2017,Langbehn2017,schindler2018,Franca2018,wang2018higher,Ezawakagome,Roy2019,Trifunovic2019,Khalaf2018,Szumniak2020,Ni2020,BiyeXie2021} 
accomplishes the field more exciting. An $n^{\rm th}$ order HOT insulator [superconductor (HOTSC)] phase is characterized by the existence of electronic [Majorana] boundary modes at their $(d-n)$-dimensional boundaries~($0 < n\le d$)~\cite{Geier2018,Zhu2018,Liu2018,Yan2018,WangWeak2018,ZengPRL2019,Zhang2019,ZhangFe2019PRL,Volpez2019,YanPRB2019,Ghorashi2019,GhorashiPRL2020,Wu2020,jelena2020HOTSC,
BitanTSC2020,SongboPRR12020,SongboPRR22020,SongboPRB2020,kheirkhah2020vortex,PlekhanovPRB2020,ApoorvTiwari2020,YanPRL2019,AhnPRL2020,luo2021higherorder2021,QWang2018,
Ghosh2021PRB,RoyPRBL2021}.

To this end, we focus on the periodically driven quantum systems, exhibiting non-trivial properties compared to their static counterparts such as dynamical localization~\cite{kayanuma08,nag14,nag15}, many-body localization~\cite{d13many,d14long,ponte15periodically}, Floquet time crystals~\cite{else16floquet,khemani16phase}, and higher harmonic generation~\cite{nag17,ikeda18} etc. In particular, anomalous boundary modes at finite quasienergy, namely $\pi$-modes, with concurrent regular $0$-modes, can be engineered by Flqouet driving~\cite{Rudner2013}. Moreover, one can architect the Floquet HOT insulators~(FHOTIs)~\cite{Bomantara2019,Nag19,YangPRL2019,Seshadri2019,Martin2019,Ghosh2020,Huang2020,HuPRL2020,YangPRR2020,Nag2020,ZhangYang2020,bhat2020equilibrium,GongPRBL2021,chaudharyphononinduced2020,JiabinYu2021,Vu2021,ghosh2021systematic,du2021weyl,ning2022tailoring} and Floquet HOT superconductors~(FHOTSCs)~\cite{PlekhanovPRR2019,BomantaraPRB2020,RWBomantaraPRR2020,RWBomantaraPRR2020,ghosh2020floquet,ghosh2020floquet2,VuPRBL2021} out of non-topological 
or lower-order topological systems. 

Till date, there exist a very few proposals, based on step-like protocol, to realize the FHOTI phase hosting both $0$- and anomalous $\pi$-mode~\cite{Huang2020,HuPRL2020,JiabinYu2021,ghosh2021systematic}. The dynamical FHOTI modes in 2D are characterized by redefining the polarization for driven systems, where the mirror symmetry plays the pivotal role~\cite{Huang2020}. Although, the hunt for such FHOTSC phases is still in its infancy~\cite{VuPRBL2021}, along with their dynamical topological characterizations. Hence, we seek  the answers for the following intriguing questions that have not been addressed so far- (a) is it possible to systematically generate the FHOTSC hosting both $0$- and the anomalous $\pi$-Majorana mode in 2D and 3D? and (b) how to characterize these $0$- and $\pi$-modes using a proper dynamical topological invariant? 

In this manuscript, we employ a periodic step-drive protocol to systematically formulate the 2D quadrupolar Floquet second-order TSC (FSOTSC) and the 3D octupolar Floquet third-order TSC~(FTOTSC), based on unconventional $d$-wave superconductor. This driving protocol allows us to realize and characterize both the $0$- and $\pi$-Majorana corner modes~(MCMs), and serves as the primary motivation of the current work. We extensively study the dynamical octupolar motion in 3D, which adds significant merit to the problem we are dealing with. 

The remainder of the article is organized as follows. We discuss the generation of anomalous Majorana modes in Sec.~\ref{Sec:II}. We topologically characterize the 2D FSOTSC and 3D FTOTSC phase using dynamical quadrupole moment and dynamical octupole moment, respectively in Sec.~\ref{Sec:III}. Finally, we summarize and conclude our paper in Sec.~\ref{Sec:IV}.

\section{Generation of anomalous Majorana modes}\label{Sec:II}
Considering the $d$-wave superconductor, we prescribe the following three-step drive protocol to foster the 2D FSOTSC and 3D FTOTSC
\begin{eqnarray}
	H_{d \rm D} (\vect{k},t)&=& J_1' h_{1, d \rm D}(\vect{k}) \  ;   \quad \quad t \in [0, T/4] \ , \nonumber \\
	&=&J_2' h_{2, d \rm D}(\vect{k}) 
	\  ; \quad \quad t \in (T/4,3T/4] \ , \nonumber \\
	&=& J_1' h_{3, d \rm D}(\vect{k}) \  ; \quad \quad t \in (3T/4,T ].
	\label{drive}
\end{eqnarray}
Here, $J'_i h_{i, d \rm D}(\vect{k})$ denotes the  Hamiltonian of the system at the $i^{\rm th}$ step in $d$-dimension~($d$D); while  $J_1'$ and $J_2'$ carry the dimensions of energy.  We define the dimensionless parameters $(J_1,J_2)=(J_1' T, J_2' T)$ where, $T$ ($\Omega=2 \pi /T$) represents the time-period  (frequency) of the drive. We  set $\hbar=c=1$. In particular, to generate a 2D FSOTSC, we choose $h_{1, 2 \rm D}(\vect{k})=h_{3, 2 \rm D}(\vect{k})= \tau_z \sigma_z$ and $h_{2, 2 \rm D}(\vect{k})=\epsilon_{2 \rm D}(\vect{k}) \tau_z \sigma_z+ \Lambda_{2 \rm D}(\vect{k}) + \Delta_{2 \rm D}(\vect{k})$; whereas, in 3D, we consider $h_{1, 3 \rm D}(\vect{k})=h_{3, 3 \rm D}(\vect{k})= \tau_z \sigma_z$ and $h_{2, 3 \rm D}(\vect{k})= \epsilon_{3 \rm D}(\vect{k}) \tau_z \sigma_z +\Lambda_{3 \rm D}(\vect{k}) + \Delta_{3 \rm D}(\vect{k})$; with $\epsilon_{2 \rm D}(\vect{k})=\left( \cos k_x + \cos k_y \right)$, $\Lambda_{2 \rm D}(\vect{k})=\sin k_x \tau_z \sigma_x s_z + \sin k_y  \tau_z \sigma_y$, $\Delta_{2 \rm D}(\vect{k})= \Delta \left( \cos k_x - \cos k_y \right) \tau_x$, $\epsilon_{3 \rm D}(\vect{k})=\left( \cos k_x + \cos k_y + \cos k_z \right)$, $\Lambda_{3 \rm D}(\vect{k})=\sin k_x \tau_z \sigma_x s_x + \sin k_y \tau_z \sigma_x s_y  + \sin k_z \tau_z   \sigma_x s_z$, and $\Delta_{3 \rm D}(\vect{k})=\Delta_1 \left( \cos k_x - \cos k_y \right)\tau_x + \Delta_2 \left( 2 \cos k_z -\cos k_x - \cos k_y \right) \tau_y$. Here, $\epsilon_{d \rm D}(\vect{k})$ and $\Lambda_{d \rm D}(\vect{k})$ encapsulate all the hoppings and spin-orbit coupling terms in $d$D, respectively. In 2D, we use the $d_{x^2-y^2}$-pairing, given by $\Delta_{2 \rm D}(\vect{k})$~\cite{Yan2018,ghosh2020floquet2} while, in 3D we incorporate mixed pairing $d_{x^2-y^2}+i d_{3z^2-r^2}$, represented by $\Delta_{3 \rm D}(\vect{k})$~\cite{RoyPRBL2021,Roydwave2019}. In the first and last step of the drive, the Hamiltonian contains only on-site term [$h_{1, d \rm D}(\vect{k})$], providing us further analytical sophistication and facilitating the topological characterizations~\cite{Rudner2013,Huang2020}. Here, both $h_{1, d \rm D}(\vect{k})$ and $h_{2, d \rm D}(\vect{k})$ respect the anti-unitary particle-hole, unitary chiral, and mirror symmetry while the last one  plays the decisive role.

\begin{figure}[]
	\centering
	\subfigure{\includegraphics[width=0.48\textwidth]{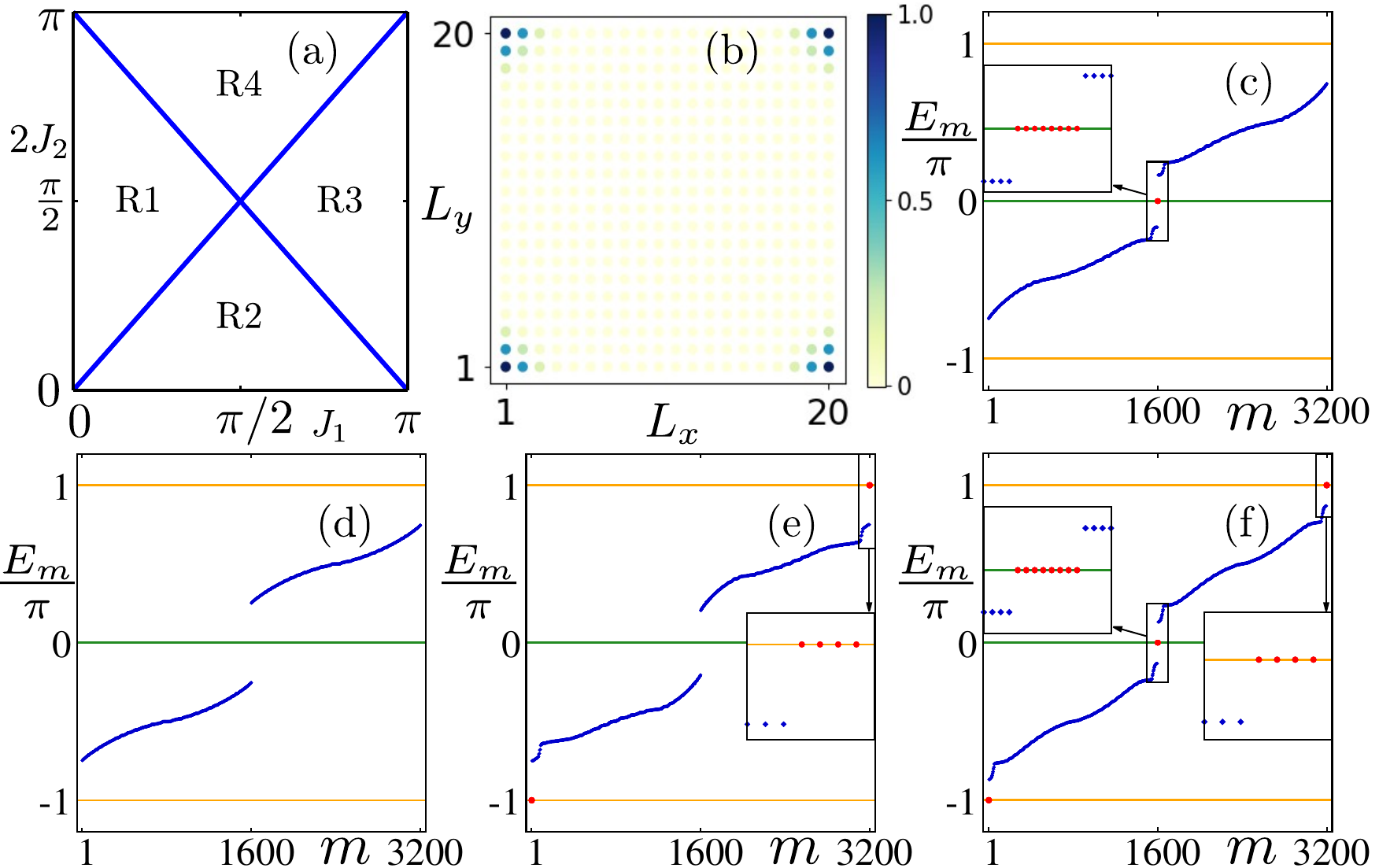}}
	\caption{(Color online) (a) We depict the phase diagram of 2D FSOTSC in $J_1-J_2$ plane (Eq.~(\ref{eq_pd})). (b) The LDOS is demonstrated for a 2D square lattice of dimension $L_x \times L_y$ 
	for $E_m=0, \pm \pi$. The quasi-energy spectra, $E_m$, computed from Eq.~(\ref{FO}), are shown as a function of the state index $m$ in panels (c), (d), (e), and (f) for R1, R2, R3, and R4, respectively. 
	We use the parameters as: $(J_1,2 J_2)=\left[ \left(\frac{\pi}{4},\frac{\pi}{2}\right), \left(\frac{\pi}{2},\frac{\pi}{4}\right),~\left(\frac{3\pi}{4},\frac{\pi}{2}\right), ~\left(\frac{\pi}{2},\frac{3 \pi}{4}\right) 
	\right]$ for R1, R2, R3, and R4, respectively. We choose $\Delta=1.0$ throughout our numerical analysis.
	}
	\label{2DSOTSC}
\end{figure}

\begin{figure}[]
	\centering
	\subfigure{\includegraphics[width=0.48\textwidth]{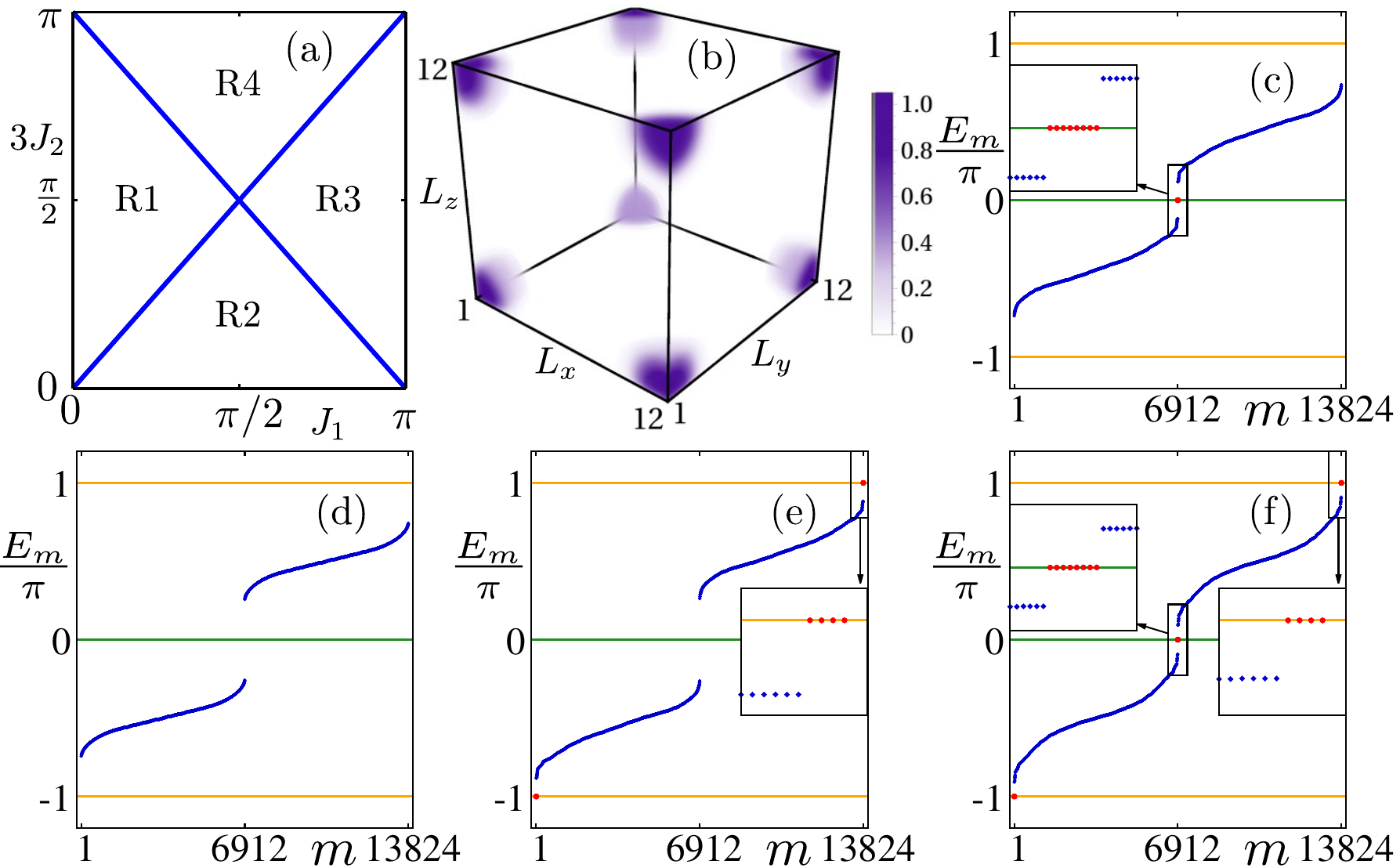}}
	\caption{(Color online) We repeat the outcome of Fig.~\ref{2DSOTSC} considering a 3D cubic lattice of dimension $L_x\times L_y \times L_z$ following Eq.~(\ref{FO}). We set $\Delta_1=\Delta_2=1.0$ 
	in the above calculation. The value of the other parameters is chosen to be the same as mentioned in Fig.~\ref{2DSOTSC}. 
	}
	\label{3DTOTSC}
\end{figure}

The Floquet operator $U(\vect{k},T)$, following the time-ordered~(TO) notation, is given as \cite{supp}
\begin{eqnarray}\label{FO}
U_{d \rm D}(\vect{k},T)&=& {\rm TO} \ \exp \left[ -i \int_{0}^{T} \ dt \ H_{d \rm D} (\vect{k},t)  \right] .
\end{eqnarray}
Using the eigenvalue equation for $U_{d \rm D}(\vect{k},T)$: $U_{d \rm D}(\vect{k},T) \ket{\Psi} =\exp \left[ {-i E(\vect{k}) } \right] \ket{\Psi}$, we obtain
\begin{eqnarray}\label{eigen}
E(\vect{k})&=& \pm \arccos \Big[ \cos \left( \alpha_{d \rm D}(\vect{k}) J_1/2 \right)  
\cos \left(\beta_{d \rm D}(\vect{k}) J_2 /2 \right) \non \\
&&- 
\sin \left( \alpha_{d \rm D}(\vect{k}) J_1 /2 \right)  
\sin \left( \beta_{d \rm D}(\vect{k}) J_2/2 \right)  \chi_{d \rm D}(\vect{k}) \Big] \ , \qquad
\end{eqnarray}
where, $\alpha_{d \rm D}(\vect{k})=\lvert h_{1, d \rm D}(\vect{k}) \rvert=\lvert h_{3, d \rm D}(\vect{k}) \rvert$, $\beta_{d \rm D}(\vect{k})=\lvert h_{2, d \rm D}(\vect{k}) \rvert$ and $\chi_{d \rm D}(\vect{k})=\frac{\epsilon_{d \rm D}(\vect{k})}{\alpha_{d \rm D}(\vect{k})\beta_{d \rm D}(\vect{k})}$. We invoke the band-gap closing across $E(\vect{k})=0,\pm \pi$ at $(k_x,k_y)=(0,0)$ or $(\pi, \pi)$ for 2D and at $(k_x,k_y,k_z)=(0,0,0)$ or $(\pi, \pi,\pi)$ for 3D to acquire the generalized topological phase boundary akin to our driving protocol in $d$-dimension as~\cite{ghosh2021systematic}
\begin{eqnarray}
\frac{d \lvert J_2 \rvert}{2}=\frac{\lvert J_1 \rvert}{2} + n \pi \ ,
\label{eq_pd}
\end{eqnarray}
where, $n \in \mathbb{Z}$. We show the topological phase diagram in the $J_1-J_2$ plane for 2D~(3D) in Fig.~\ref{2DSOTSC}(a)~(Fig.~\ref{3DTOTSC}(a)). The phase diagram can be divided into four segments- region-1~(R1) with only $0$-modes, region-2~(R2) without any modes, region-3 supporting only $\pi$-modes, and region-4~(R4) allowing both $0$ and $\pi$-MCMs to coexist.

Having perceived the problem analytically, we anchor our findings with numerical results. The 2D FSOTSC and 3D FTOTSC can be identified by the presence of zero-dimensional~(0D) MCMs~\cite{Ghosh2021PRB}; these are computed for both regular and anomalous modes while diagonalizing the Floquet operator (Eq.~(\ref{FO})) with open boundary condition~(OBC) in all directions. The corresponding local density of states (LDOS) of MCMs is shown in Fig.~\ref{2DSOTSC}~(b) and Fig.~\ref{3DTOTSC}~(b), respectively for 2D square and 3D cubic lattice. We portray the quasienergy spectra 
($E_m$) for R1 (eight $0$-MCMs), R2 (no MCMs), R3 (four MCMs each at $E_m=\pm \pi$), and R4 (eight MCMs at $E_m=0$ and four MCMs each at $E_m=\pm \pi$) in Fig.~\ref{2DSOTSC}~(c), (d), (e), and (f) [Fig.~\ref{3DTOTSC}~(c), (d), (e), and (f)], respectively, considering 2D [3D] system. Note that in 3D, both the surface and the hinge mode become gapped. Generation of these anomalous dynamical MCMs via our three-step driving protocol is one of the main results of this article.

\section{Topological characterization of FHOTSC phase}\label{Sec:III}
 For the anomalous Floquet phase, the main challenge is to topologically characterize both the $0$- and $\pi$-MCMs distinctively. We first pursue the appropriate Wannier sector polarization for 2D FSOTSC and 3D FTOTSC, employing nested Wilson loop techniques~\cite{benalcazarprb2017,Ni2020,supp}, from the Floquet operator $U_{d \rm D}(\vect{k},T)$. For 2D FSOTSC [3D FTOTSC], the average first-order [second-order] nested Wannier polarization for $\mu'$ [$\mu''$]-$^{\rm th}$ sector $\langle \nu_{y,\rm Flq, \mu'}^{\pm \nu_{x}} \rangle $ [$\langle \nu_{z,\rm Flq, \mu''}^{\pm \nu_{y }^{\pm \nu_{x}}} \rangle$]~\cite{supp} exhibits a quantized value of $0.5$, when the system is in the regime R1 and R3. However, the same is unable to ascertain one whether the modes are lying at $0$ or $\pi$-gap. These nested polarizations reduce to $0$ for both the trivial phase in R2 and the anomalous phase hosting both $0$ and $\pi$-mode in R4.

Inadequacy of the topological invariant, computed from the \textit{quasi-static} Floquet operator, motivates us to hunt for a dynamical topological invariant (both in 2D and 3D) that cannot only extricate 
R2 from R4 but also unmistakably yields distinct signatures of $0$ and $\pi$-modes. We consider the full time evolution operator $U_{d \rm D}(\vect{k},t)$, embodying an \textit{anomalous} periodized 
part $U_{d\rm D,\epsilon}(\vect{k},t)$ and a \textit{normal} quasi-static part $\left[U(\vect{k},T)\right]^{t/T}_\epsilon$, such that~\cite{Rudner2013,Huang2020}
\begin{eqnarray}\label{time-evolution}
U_{d \rm D}(\vect{k},t)=U_{d \rm D,\epsilon}(\vect{k},t)\left[U_{d \rm D}(\vect{k},T)\right]^{t/T}_\epsilon \ ,
\end{eqnarray}
here, the subscript $\epsilon$ denotes the $0$ and $\pi$-gap and enables us to keep track of the origin of the MCMs in these quasi-energies. We use the periodized evolution operator~(PEO) $U_{d \rm D,\epsilon}(\vect{k},t)=U_{d \rm D}(\vect{k},t)\left[U_{d \rm D}(\vect{k},T)\right]^{-t/T}_\epsilon$ to calculate the pertinent topological invariant. However, unlike $U_{d \rm D}(\vect{k},T)$ the PEO does not possess any conventional band physics and can be gapless at certain time-instants~\cite{Rudner2013,Huang2020}.

In order to capture the non-triviality of the anomalous Floquet modes, we define the dynamical mean polarization as relative motion of a particle at two instants~\cite{Huang2020,supp}
\begin{equation}
 \hat{\bar{x}}(t)=\frac{\left[\hat{x}(t)+\hat{x}(0)\right]}{2} \ ,
\end{equation}
here, $\hat{x}(0)=\hat{x}=\sum_{im} \hat{c}_{im}^\dagger \ket{0} e^{-i \Delta_x x_i} \bra{0} \hat{c}_{im}$ exemplify the static polarization~\cite{RestaPRL1998} with $\Delta_i=2 \pi / L_i$, $\hat{x}(t) = U^\dagger_{d \rm D,\epsilon}(\vect{k},t) \ \hat{x} \ U_{d \rm D,\epsilon}(\vect{k},t)$ and $\hat{c}$'s being the quasiparticle creation operators. The eigenvalue of $\hat{\bar{x}}(t)$ is related to the dynamical Wilson loop operator $W_{x,\epsilon,\vect{k}}(t)$ in the following way: $\left(\hat{\bar{x}}(t)\right)^{L_x}$$ = \sum_{\vect{k}mn} \hat{c}_{\vect{k}m}^\dagger \ket{0} \left[W_{x,\epsilon,\vect{k}}(t)\right]_{mn} \bra{0} \hat{c}_{\vect{k}n}$. One can find $W_{x,\epsilon,\vect{k}}(t)= $ $ Q_{x,\epsilon,\vect{k}+(L_x-1)\Delta_x \vect{e}_x}(t) \cdots Q_{x,\epsilon,\vect{k}+\Delta_x \vect{e}_x}(t) $ $Q_{x,\epsilon,\vect{k}}(t)$;~with $Q_{p,\epsilon,\vect{k}}(t)=\frac{\mathbb{I} + U^\dagger_{d \rm D,\epsilon}(\vect{k}+\Delta_p\vect{e}_p,t) U_{d \rm D,\epsilon}(\vect{k},t)}{2}$, and the unit vector along $p^{\rm th}$ direction is represented by $\vect{e}_p$. From the eigenvalue equation for $W_{x,\epsilon,\vect{k}}(t)$ : $W_{x,\epsilon,\vect{k}}(t) \ket{\nu_{x,\epsilon , \mu}(\vect{k},t)} = e^{-2 \pi i \nu_{x,\epsilon, \mu}(k_{j \neq x},t)}\ket{\nu_{x,\epsilon , \mu}(\vect{k},t)}$, one obtains the dynamical first-order branches $\nu_{x,\epsilon, \mu}(k_{j \neq x},t)$. Here, $\nu_{x,\epsilon, \mu}(k_{j \neq x},t)$ refers to a relative motion of a particle along $x$-direction with respect to $x_i$  during $t \in \left[0,t\right]$~\cite{Huang2020}. This dynamical first-order branches can  characterize anomalous Floquet first-order topological phase. In order to conceive the higher-order moments, one needs to incorporate the nested structure while constructing the Wilson loop, as executed for the static systems with appropriate $Q_{p,\vect{k}}$~\cite{benalcazarprb2017}.  

\begin{figure}[]
	\centering
	\subfigure{\includegraphics[width=0.49\textwidth]{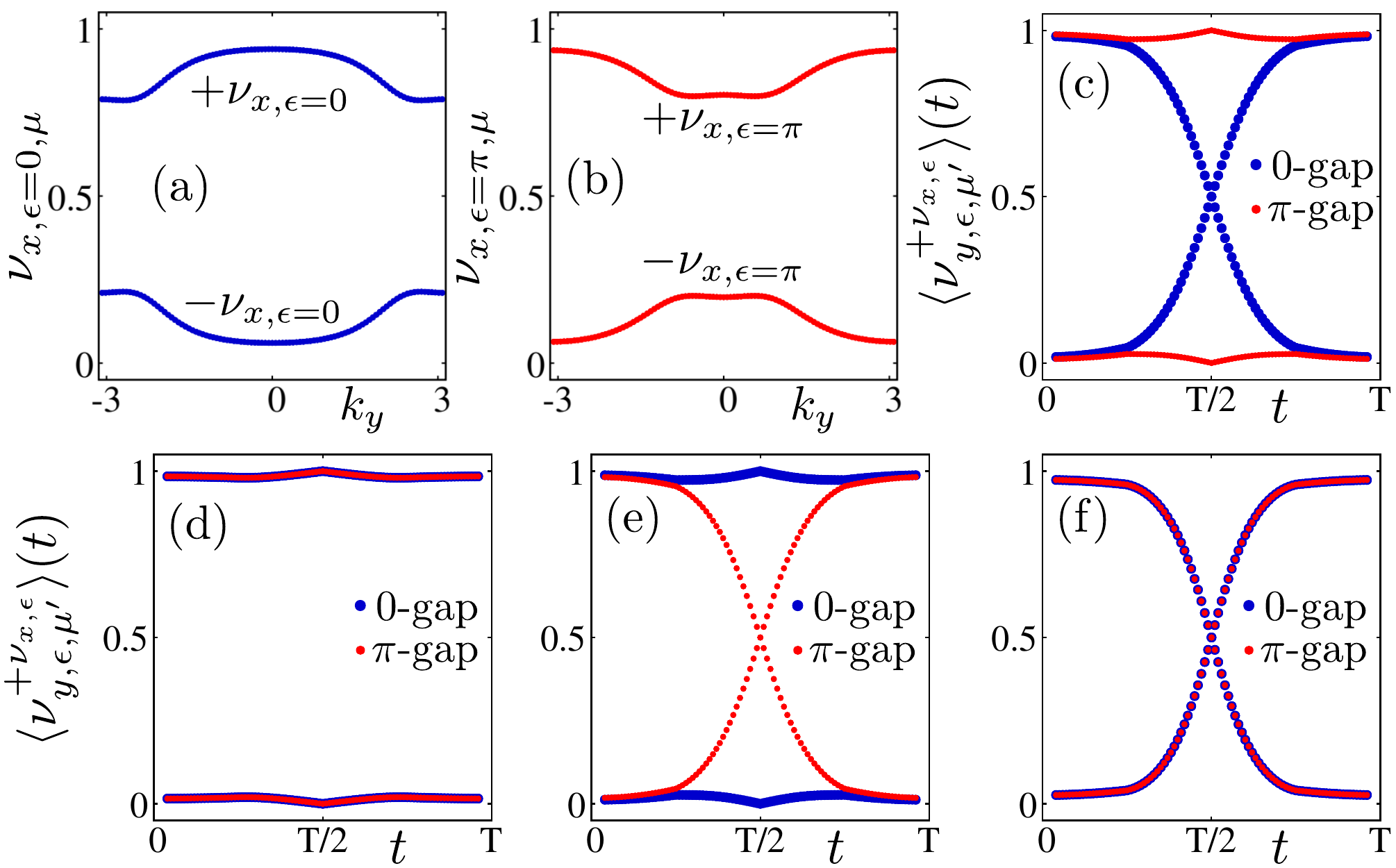}}
	\caption{(Color online) The gapped dynamical polarization branches $\nu_{x,\epsilon=0, \mu}$  and $\nu_{x,\epsilon=\pi, \mu}$, arising from $0$ and $\pi$-gap, are respectively shown as a function of  
	$k_y$ in (a) and (b), at time $t=\frac{T}{2}$, while the system is in R4. The average quadrupolar motion $\langle \nu_{y,\epsilon , \mu'}^{+\nu_{x,\epsilon}} \rangle (t)$ for R1, R2, R3, and R4 (see 
	Fig.~\ref{2DSOTSC}) are depicted in panels (c), (d), (e), and (f), respectively, as a function of time $t$, manifesting gapless crossing between opposite branches. Here, blue and red dots represents 
	$\langle \nu_{y, \epsilon , \mu'}^{+\nu_{x,\epsilon}} \rangle (t)$ arising from $0$ and $\pi$-gap, respectively. See text for discussion. 
	}
	\label{quadrupole}
\end{figure}

To accommodate the FHOTSC phase, the eigenvalues $\nu_{x,\epsilon, \mu}(k_{j \neq x},t)$ remain gapped during the full cycle $t\in\left[0,T\right]$ and can be grouped into two separable sets $\pm \nu_{x,\epsilon}$. The dynamical second-order polarization is computed by evaluating a relative motion of particle along $y$-direction by projecting onto the set $\pm \nu_{x,\epsilon}$ with projector $P_{\pm \nu_{x, \epsilon}} (t)$ \cite{supp}: $\hat{\bar{y}}^{\pm \nu_{x,\epsilon}} (t)= P_{\pm \nu_{x, \epsilon}} (t)  \hat{\bar{y}} (t)  P_{\pm \nu_{x, \epsilon}} (t)$. Similar to the earlier case, we can obtain dynamical 
first-order nested Wilson loop operator $W_{y,\epsilon,\vect{k}}^{\pm \nu_{x,\epsilon}} (t)$ from $\hat{\bar{y}}^{\pm \nu_{x,\epsilon}} (t)$: $\left(\hat{\bar{y}}^{\pm \nu_{x,\epsilon}} (t)\right)^{L_y}=\sum_{\vect{k},\mu_1,\mu_2 \in \pm \nu_{x, \epsilon}} \gamma_{\vect{k} \epsilon \mu_1 }^\dagger(t) \ket{0} \left[W_{y,\epsilon,\vect{k}}^{\pm \nu_{x,\epsilon}} (t) \right]_{\mu_1\mu_2} \bra{0} \gamma_{\vect{k} \epsilon \mu_2 } (t)$ where $\gamma$'s are constituted from $\ket{\nu_{x,\epsilon , \mu}(\vect{k},t)}$ according the projection rule~\cite{supp}. This leads to    
$W_{y,\epsilon,\vect{k}}^{\pm \nu_{x,\epsilon}} (t)= Q_{y,\epsilon,\vect{k}+(L_y-1)\Delta_y \vect{e}_y}^{\pm \nu_{x,\epsilon}}(t) \cdots Q_{y,\epsilon,\vect{k}+\Delta_y \vect{e}_y}^{\pm \nu_{x,\epsilon}}(t) Q_{y,\epsilon,\vect{k}}^{\pm \nu_{x,\epsilon}}(t)$ with $\left[Q_{y,\epsilon,\vect{k}}^{\pm \nu_{x, \epsilon}} (t)\right]_{\mu_1 \mu_2} = \sum_{m n} \left[\nu_{x,\epsilon , \mu_1} (\vect{k}+\Delta_y \vect{e}_y,t) \right]^*_m $ $ \left[Q_{y,\epsilon,\vect{k}}(t)\right]_{mn} \left[\nu_{x,\epsilon , \mu_2} (\vect{k},t) \right]_n$. 
The dynamical second-order quadrupolar branches $\nu_{y,\epsilon, \mu'}^{\pm \nu_{x,\epsilon}}(k_{j \neq y},t)$ can be obtained from the eigenvalue equation : $W_{y,\epsilon,\vect{k}}^{\pm \nu_{x,\epsilon}} (t) \ket{\nu_{y,\epsilon , \mu'}^{\pm \nu_{x,\epsilon}}(\vect{k},t)} = e^{-2 \pi i \nu_{y,\epsilon, \mu'}^{\pm \nu_{x,\epsilon}}(k_{j \neq y},t)}\ket{\nu_{y,\epsilon , \mu'}^{\pm \nu_{x,\epsilon}}(\vect{k},t)}$. This quadrupolar branches can topologically characterize the anomalous 2D FSOTSC, which we illustrate in Fig.~\ref{quadrupole}. 

\begin{figure}[]
	\centering
	\subfigure{\includegraphics[width=0.5\textwidth]{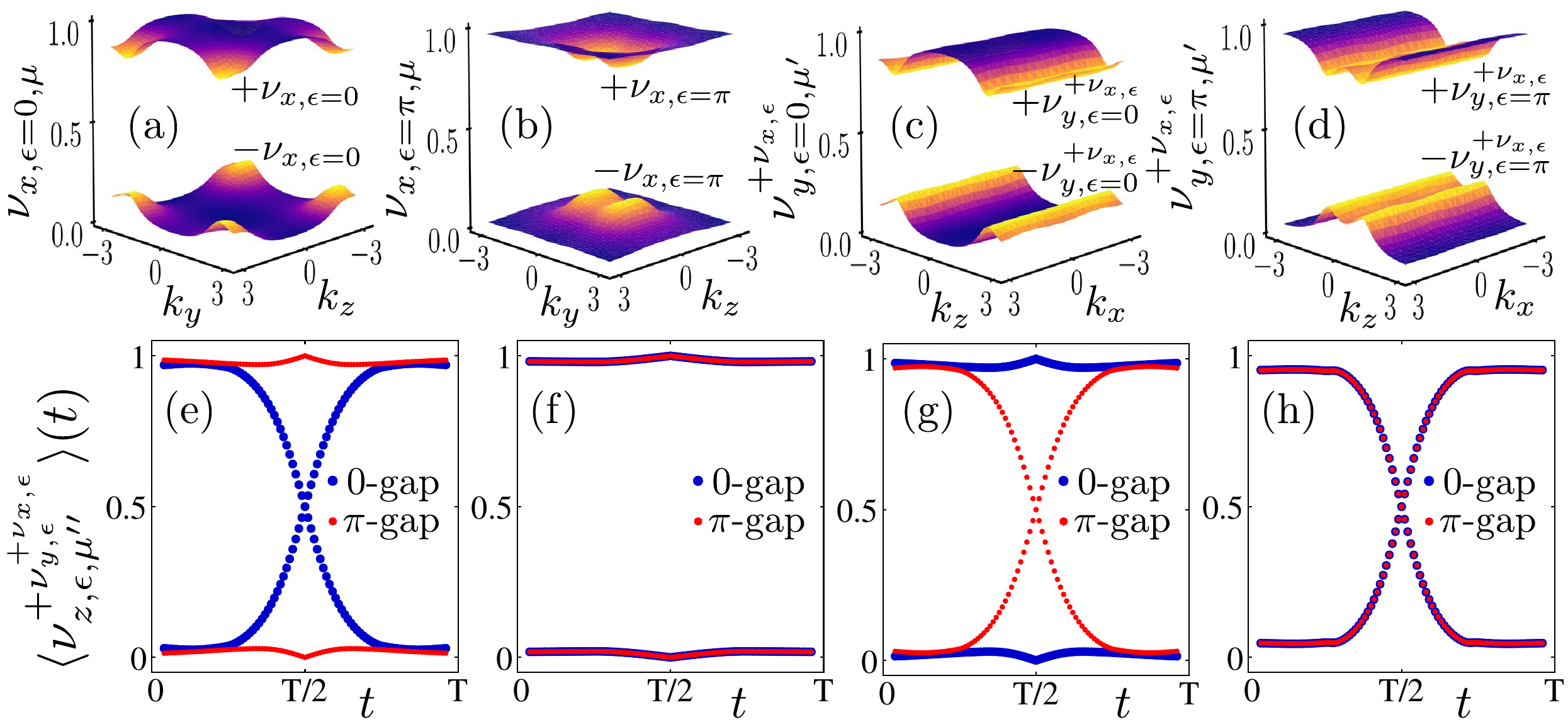}}
	\caption{(Color online) We repeat Figs.~\ref{quadrupole} (a), and (b) for the 3D cubic lattice 
	as a function of $k_z,~k_x$ and depict in panels (a) and (b), respectively. The gapped dynamical quadrupolar branches $\nu_{y,\epsilon=0 , \mu'}^{+\nu_{x,\epsilon}}$ and $\nu_{y,\epsilon=\pi , \mu'}^{+
	\nu_{x,\epsilon}}$  are shown in panels (c) and (d), respectively, at $t=\frac{T}{2}$ in R4. The average octupolar motion $\langle \nu_{z,\epsilon, \mu''}^{+ \nu_{y,\epsilon }^{+ \nu_{x,\epsilon}}} \rangle (t)
	$ for R1, R2, R3, and R4 are shown in panels (c), (d), (e), and (f), respectively, indicating the gapless crossing between opposite branches. Here, blue and red dots represent $\langle \nu_{z,\epsilon,  
	\mu''}^{+ \nu_{y,\epsilon }^{+ \nu_{x,\epsilon}}} \rangle (t)$ arising from $0$ and $\pi$-gap, respectively and are discussed in the text.
	}
	\label{octupole}
\end{figure}

Proceeding further, the octupolar phase guarantees gapped quadrupolar branches during time $t \in \left[0,T\right]$ and thus, can be grouped into two dissociable sets $\pm\nu_{y,\epsilon}^{\pm \nu_{x,\epsilon}}$. The dynamical third-order polarization can be portrayed as the relative motion of particle along the remaining $z$-direction using the projector $P_{\pm \nu_{y,\epsilon }^{\pm \nu_{x,\epsilon}}} (t)$ onto  $\pm\nu_{y,\epsilon}^{\pm \nu_{x,\epsilon}}$: $\hat{\bar{z}}^{\pm \nu_{y,\epsilon }^{\pm \nu_{x,\epsilon}}} (t)= P_{\pm \nu_{y,\epsilon }^{\pm \nu_{x,\epsilon}}} (t) \ \hat{\bar{z}} (t) \ P_{\pm \nu_{y,\epsilon }^{\pm \nu_{x,\epsilon}}} (t)$~\cite{supp}. Following the similar line of argument, the dynamical second-order nested Wilson loop operator is found to be $\left(\hat{\bar{z}}^{\pm \nu_{y,\epsilon }^{\pm \nu_{x,\epsilon}}}(t)\right)^{L_z}=\sum_{\vect{k},\mu'_1,\mu'_2 \in \pm \nu_{y,\epsilon }^{\pm \nu_{x,\epsilon}}} \eta_{\vect{k} \epsilon \mu'_1 }^\dagger(t) \ket{0} \left[W_{z,\epsilon,\vect{k}}^{\pm \nu_{y,\epsilon }^{\pm \nu_{x,\epsilon}}} (t)\right]_{\mu'_1 \mu'_2} \bra{0} \eta_{\vect{k} \epsilon \mu'_2 } (t)$ where $\eta$'s are comprised from $\ket{\nu_{x,\epsilon , \mu}(\vect{k},t)}$ and $\ket{\nu_{y,\epsilon , \mu'}^{\pm \nu_{x,\epsilon}}(\vect{k},t)}$~\cite{supp}. We, therefore, obtain $W_{z,\epsilon,\vect{k}}^{\pm \nu_{y,\epsilon }^{\pm \nu_{x,\epsilon}}} (t)= Q_{z,\epsilon,\vect{k}+(L_z-1)\Delta_z \vect{e}_z}^{\pm \nu_{y,\epsilon }^{\pm \nu_{x,\epsilon}}}(t) \cdots Q_{z,\epsilon,\vect{k}+\Delta_z \vect{e}_z}^{\pm \nu_{y,\epsilon }^{\pm \nu_{x,\epsilon}}}(t) Q_{z,\epsilon,\vect{k}}^{\pm \nu_{y,\epsilon }^{\pm \nu_{x,\epsilon}}}(t)$, with $\left[Q_{z,\epsilon,\vect{k}}^{\pm \nu_{y,\epsilon }^{\pm \nu_{x,\epsilon}}} (t)\right]_{\mu'_1 \mu'_2} =\sum_{\substack{m n \mu_1 \mu_2}} \left[ \nu_{y,\epsilon , \mu'_1}^{\pm \nu_{x,\epsilon}}(\vect{k}+\Delta_z \vect{e}_z,t) \right]_{\mu_1}^*$ $\left[ \nu_{x,\epsilon , \mu_1}(\vect{k}+\Delta_z \vect{e}_z,t) \right]_m^* \left[Q_{z,\epsilon,\vect{k}}(t)\right]_{mn}\left[ \nu_{x,\epsilon , \mu_2}(\vect{k},t) \right]_n$ $\left[ \nu_{y,\epsilon , \mu'_2}^{\pm \nu_{x,\epsilon}}(\vect{k},t) \right]_{\mu_2}$. From the eigenvalue of $W_{z,\epsilon,\vect{k}}^{\pm \nu_{y,\epsilon }^{\pm \nu_{x,\epsilon}}} (t)$, we procure the dynamical third-order octupolar branch as $\nu_{z,\epsilon, \mu''}^{\pm \nu_{y,\epsilon }^{\pm \nu_{x,\epsilon}}}(k_{j \neq z},t)$ that is adopted to topologically characterize the 3D FTOTSC as demonstrated in Fig.~\ref{octupole}.

\subsection{ Dynamical quadrupole moment} 
Limited to 2D, the mirror symmetry $\mathcal{M}_x$ enforces $\nu_{x,\epsilon, \mu}(k_{y},t)$ to appear in pairs: $\nu_{x,\epsilon, \mu_1}(k_{y},t)=-\nu_{x,\epsilon, \mu_2}(k_{y},t)$, with $\mu_1 \in +\nu_{x,\epsilon, \mu}$ and $\mu_2 \in -\nu_{x,\epsilon, \mu}$. Moreover, $\mathcal{M}_y$ compels $\nu_{x,\epsilon, \mu}(k_{y},t)=\nu_{x,\epsilon, \mu}(-k_{y},t)$ within each branch $\mu$~\cite{Huang2020}. We show these behavior in Fig.~\ref{quadrupole}~(a) and (b) for $0$ and $\pi$-gap at $t=\frac{T}{2}$, respectively, when the system is in R4. $\mathcal{M}_x$ imposes the quadrupolar branches, derived from opposite first-order branches, to be the same \ie $\nu_{y,\epsilon, \mu'}^{+ \nu_{x,\epsilon}}(k_x,t)=\nu_{y,\epsilon, \mu'}^{- \nu_{x,\epsilon}}(k_x,t)$ and $\mathcal{M}_y$ causes the quadrupolar branches to appear in pairs: $\nu_{y,\epsilon, \mu'_1}^{+ \nu_{x,\epsilon}}(k_x,t)=-\nu_{y,\epsilon, \mu'_2}^{+ \nu_{x,\epsilon}}(k_x,t)$~\cite{Huang2020}.
It is evident from the above discussion that the mirror symmetries do not impose any constraints on the quantization of the quadrupolar branches at any time-instant $t$ unlike the static case where 
$\nu$'s are allowed to take values either $0$ or $1/2$ (mod $1$)~\cite{benalcazarprb2017}. 

The average quadrupolar motion $\langle \nu_{y,\epsilon , \mu'}^{+ \nu_{x,\epsilon}} \rangle (t)=\frac{1}{L_x} \sum_{k_x} \nu_{y,\epsilon , \mu'}^{+ \nu_{x,\epsilon}} (k_x,t)$, however, plays a paramount role in the understandings of the topological Floquet modes. Now, $U_{2 \rm D, \epsilon}(0)=U_{2 \rm D, \epsilon}(T)=\mathbb{I}$, necessitates the particles to undergo a round trip during the time-interval $t \in \left[0,T\right]$ and enforces $\langle \nu_{y,\epsilon , \mu'}^{+ \nu_{x,\epsilon}} \rangle (t=0)=\langle \nu_{y,\epsilon , \mu'}^{+ \nu_{x,\epsilon}} \rangle (t=T)=0~({\rm mod}~1)$ to be the fixed-points. For a topologically trivial phase, $\langle \nu_{y,\epsilon , \mu'}^{+ \nu_{x,\epsilon}} \rangle (0)$ and $\langle \nu_{y,\epsilon , \mu'}^{+ \nu_{x,\epsilon}} \rangle (T)$ are adiabatically connected without any gap closing between two branches $\forall t \in \left[0,T\right]$ (see Fig.~\ref{quadrupole}~(c) $\pi$-gap, (d) both gaps, and (e) $0$-gap). The presence of MCMs (see Fig.~\ref{2DSOTSC}) in the gap $\epsilon$, obstructs the motion of $\langle \nu_{y,\epsilon , \mu'}^{+ \nu_{x,\epsilon}} \rangle (t)$ in the time-interval $t \in \left[0,T\right]$ and two branches cross each other at $\frac{1}{2}~({\rm mod}~1)$ at $t =\frac{T}{2}$ (see Fig.~\ref{quadrupole}~(c) $0$-gap, (e) $\pi$-gap, and (f) both $0$ and $\pi$-gap). We thus obtain the quantization of the dynamical quadrupole moment $Q_{\epsilon}^{+ \nu_{x,\epsilon}}=\int_{0}^{T} dt \partial_t \langle \nu_{y,\epsilon , \mu'}^{+ \nu_{x,\epsilon}} \rangle (t)=[0]~1~({\rm mod}~1)$, for [trivial] topological case, giving rise to a $\mathbb{Z}_2$ classification. Thus, the generalization of dynamical quadrupole moment for the 2D FSOTSC (with eight-band model) is another important result of this manuscript.
\subsection{Dynamical octupole moment} 
 In 3D, $\mathcal{M}_x$ seeks $\nu_{x,\epsilon, \mu_1}(k_y,k_z,t)=-\nu_{x,\epsilon, \mu_2}(k_y,k_z,t)$, with $\mu_1 \in +\nu_{x,\epsilon, \mu}$ and $\mu_2 \in -\nu_{x,\epsilon, \mu}$. While, $\mathcal{M}_y$ and $\mathcal{M}_z$ set the shape of the branch such that  $\nu_{x,\epsilon, \mu}(k_y,k_z,t)=\nu_{x,\epsilon, \mu}(-k_y,k_z,t)$ and $\nu_{x,\epsilon, \mu}(k_y,k_z,t)=\nu_{x,\epsilon, \mu}(k_y,-k_z,t)$ within each branch $\mu$. We demonstrate the dynamical first-order branch in Figs.~\ref{octupole}~(a) and (b) while the system is in R4 at 
$t=\frac{T}{2}$ for $0$ and $\pi$-gap, respectively. 
Here, $\mathcal{M}_x$ invokes second-order branches calculated from opposite first-order branches $\pm \nu_{x,\epsilon, \mu}(k_y,k_z,t)$ to be identical, $\mathcal{M}_y$ requires $\nu_{y,\epsilon, \mu'_1}^{+ \nu_{x,\epsilon}}(k_z,k_x,t)=-\nu_{y,\epsilon, \mu'_2}^{+ \nu_{x,\epsilon}}(k_z,k_x,t)$ and $\mathcal{M}_z$ enforces $\nu_{y,\epsilon, \mu'}^{+ \nu_{x,\epsilon}}(k_z,k_x,t)=\nu_{y,\epsilon, \mu'}^{+ \nu_{x,\epsilon}}(-k_z,k_x,t)$. We depict the dynamical second-order branch in Figs.~\ref{octupole}~(c) and (d) for $0$ and $\pi$-gap, respectively, at $t=\frac{T}{2}$, while the system is in R4. Akin to the first-order branches, the quadrupolar branches also exhibit a finite gap and sets the stage for the calculation of the third-order (octupolar) dynamical branch. Here, $\mathcal{M}_x$ and $\mathcal{M}_y$ ensure octupolar branch calculated from different quadrupolar branches to remain same and $\mathcal{M}_z$ compels the octupolar branch to appear in pairs: $\nu_{z,\epsilon, \mu''_1}^{+ \nu_{y,\epsilon }^{+ \nu_{x,\epsilon}}}(k_x,k_y,t)=-\nu_{z,\epsilon, \mu''_2}^{+ \nu_{y,\epsilon }^{+ \nu_{x,\epsilon}}}(k_x,k_y,t)$. 

Following the 2D case, we introduce the average octupolar motion as $\langle \nu_{z,\epsilon, \mu''}^{+ \nu_{y,\epsilon }^{+ \nu_{x,\epsilon}}} \rangle (t)=\frac{1}{L_x L_y} \sum_{k_x k_y} \nu_{z,\epsilon, \mu''}^{+ \nu_{y,\epsilon }^{+ \nu_{x,\epsilon}}} (k_x,k_y,t)$. For the trivial case, $\langle \nu_{z,\epsilon, \mu''}^{+ \nu_{y,\epsilon }^{+ \nu_{x,\epsilon}}} \rangle (t:0\to T)$ winds back to the original value (mod $1$) without experiencing any gap closing among the branches (see Fig.~\ref{octupole}~(e) $\pi$-gap, (f) both $0$ and $\pi$-gap, and (g) $0$-gap). The topologically non-trivial situation refers to a gap 
closing of two different octupolar branches at $\frac{1}{2}~({\rm mod}~1)$ when $\langle \nu_{z,\epsilon, \mu''}^{+ \nu_{y,\epsilon }^{+ \nu_{x,\epsilon}}} \rangle (t \rightarrow 0)=0~(1)$ evolves to $\langle \nu_{z,\epsilon, \mu''}^{+ \nu_{y,\epsilon }^{+ \nu_{x,\epsilon}}} \rangle (t \rightarrow T)=1~(0)$ as depicted in Fig.~\ref{octupole}~(e) $0$-gap, (g) $\pi$-gap, (h) both $0$ and $\pi$-gap. Hence, the notion of $\mathbb{Z}_2$ invariant works for the 3D octupole moment similar to the 2D quadrupolar moment: 
$O_{\epsilon}^{+ \nu_{y,\epsilon }^{+ \nu_{x,\epsilon}}}=\int_{0}^{T} dt \partial_t \langle \nu_{z,\epsilon, \mu''}^{+ \nu_{y,\epsilon }^{+ \nu_{x,\epsilon}}} \rangle (t)=[0]~1~({\rm mod}~1)$, for [trivial] topological case. We emphasize that the topological characterization of anomalous MCMs $(0-\pi)$ via the dynamical octupole moment in the FTOTSC phase is the prime result of this article. 

\section{Summary and conclusions}\label{Sec:IV}
 To summarize, in this article, we prescribe a step-drive protocol to dynamically construct 2D FSOTSC and 3D FTOTSC hosting 0D MCMs. Exploiting 
the phase diagrams, we illustrate the emergence of both regular $0$ and anomalous $\pi$-MCMs separately and simultaneously. We circumvent the elusive affair of complete topological characterization for available dynamical phases by analyzing PEO in both 2D and 3D. This allows us 
to tie up the dynamical quadrupole and octupole moments with $\mathbb{Z}_2$ classifications and enable us to topologically characterize the FHOTSC phases. 
Along this direction, the stability of these dynamic phases in presence of strong disorder might also be an intriguing future direction.

Note that, the Majorana-based qubit architectures have been studied for FSOTSC, hosting both $0$ and anomalous $\pi$-modes, in the context of fault-tolerant quantum computing~\cite{BomantaraPRB2020}. We believe that the spatially separated MCMs in FTOTSC, observed for 
the present case, can become potentially useful for futher extension of the quantum gate operations in 3D. The three-step periodic driving protocol implemented here is found to be very covenient for the model based studies~\cite{Huang2020}. Given the experimental advancement in Floquet driving~\cite{Wang453,Experiment2016,maczewsky2017observation} and HOT phases~\cite{serra2018observation,schindler2018higher,xue2019acoustic}, our proposal carries 
possible implication of practical relevance~\cite{experimentFloquetHOTI}. However, from the experimental viewpoint, the engineering of HOT phases employing laser drive/light fields can be 
more realistic that we leave for future investigation and will be presented elsewhere. We believe that the present scheme of topological characterization would work for the continuous time driving.
\subsection*{Acknowledgments}
A.K.G. and A.S. acknowledge SAMKHYA: High-Performance Computing Facility provided by Institute of Physics, Bhubaneswar, for numerical computations.

\bibliography{bibfile}{}
\clearpage
\begin{onecolumngrid}
	\begin{center}
		{\fontsize{12}{12}\selectfont
			\textbf{Supplemental Material for ``Dynamical construction of quadrupolar and octupolar topological superconductors''\\[5mm]}}
		{\normalsize Arnob Kumar Ghosh\orcidA{},$^{1,2}$ Tanay Nag\orcidB{},$^{3}$ and Arijit Saha\orcidC{}$^{1,2}$\\[1mm]}
		{\small $^1$\textit{Institute of Physics, Sachivalaya Marg, Bhubaneswar-751005, India}\\[0.5mm]}
		{\small $^2$\textit{Homi Bhabha National Institute, Training School Complex, Anushakti Nagar, Mumbai 400094, India}\\[0.5mm]}
		{\small $^3$\textit{Institut f\"ur Theorie der Statistischen Physik, RWTH Aachen University, 52056 Aachen, Germany}\\[0.5mm]}
		{}
	\end{center}
	\normalsize
	\begin{center}
		\parbox{16cm}{In this supplemental material, we provide the detailed construction of the Floquet operator in Sec.~\ref{Sec:S1}. In Sec.~\ref{Sec:S2}, we present the periodized evolution operator. Sec.~\ref{Sec:S3} is devoted to the elaborated formalism for computing the dynamical multipole moments employed in the main text to characterize the two-dimensional (2D) Floquet second-order topological superconductor~(FSOTSC) and three-dimensional (3D) Floquet third-order topological superconductor~(FTOTSC). In Sec.~\ref{Sec:S4}, we provide the calculation for quasi-static multipole moments. Finally, in Sec.~\ref{Sec:S5}, we discuss the appearance of FSOTSC in 3D based on our driving protocol mentioned in the main text.}
	\end{center}
	\newcounter{defcounter}
	\setcounter{defcounter}{0}
	\setcounter{equation}{0}
	\renewcommand{\theequation}{S\arabic{equation}}
	\setcounter{figure}{0}
	\renewcommand{\thefigure}{S\arabic{figure}}
	\setcounter{page}{1}
	\pagenumbering{roman}
	\setcounter{section}{0}
	\renewcommand{\thesection}{S\arabic{section}}

	\section{Construction of Floquet operator} \label{Sec:S1}
	Following the step-drive protocol introduced in the main text, the evolution operator in the time interval $t \in \Big[0, \frac{T}{4} \Big]$ can be written as
	\begin{eqnarray}\label{U1}
	U_{d \rm D}(\vect{k},t)&=& \exp \left( -i J'_1 h_{1,d \rm D} (\vect{k}) t \right) \non \\
	&=& \cos \left( J_1' \alpha_{d \rm D}(\vect{k}) t \right) \mathbb{I} 
	- i \sin \left( J_1' \alpha_{d \rm D}(\vect{k}) t \right) \frac{h_{1,d \rm D} (\vect{k})}{\alpha_{d \rm D}(\vect{k})} \ .
	\end{eqnarray}
	In the interval $t \in \Big(\frac{T}{4}, \frac{3T}{4} \Big]$, the evolution operator becomes 
	\begin{eqnarray}\label{U2}
	U_{d \rm D}(\vect{k},t)&=& \exp \left( -i J'_2 h_{2,d \rm D} (\vect{k}) \left(t-\frac{T}{4}\right) \right) \exp \left( -i J'_1 h_{1,d \rm D} (\vect{k}) \frac{T}{4} \right) \non \\
	&=&\left[ \cos \left( J_2' \beta_{d \rm D}(\vect{k}) \left(t-\frac{T}{4}\right) \right) \mathbb{I} 
	- i \sin \left( J_2' \beta_{d \rm D}(\vect{k}) \left(t-\frac{T}{4}\right) \right) \frac{h_{2,d \rm D} (\vect{k})}{\beta_{d \rm D}(\vect{k})} \right] \times \non \\
	&& \hspace{4cm}\left[ \cos \left( J_1' \alpha_{d \rm D}(\vect{k}) \frac{T}{4} \right) \mathbb{I} 
	- i \sin \left( J_1' \alpha_{d \rm D}(\vect{k}) \frac{T}{4} \right) \frac{h_{1,d \rm D} (\vect{k})}{\alpha_{d \rm D}(\vect{k})} \right] \ .
	\end{eqnarray}
	Whereas, in the final step \ie $t \in \Big(\frac{3T}{4}, T \Big]$, the evolution operator reads
	\begin{eqnarray}\label{U3}
	U_{d \rm D}(\vect{k},t)&=& \exp \left( -i J'_1 h_{1,d \rm D} (\vect{k}) \left(t-\frac{3T}{4}\right) \right) \exp \left( -i J'_2 h_{2,d \rm D} (\vect{k}) \frac{T}{2} \right) \exp \left( -i J'_1 h_{1,d \rm D} (\vect{k}) \frac{T}{4} \right) \non \\
	&=&\left[ \cos \left( J_1' \alpha_{d \rm D}(\vect{k}) \left(t-\frac{3T}{4}\right) \right) \mathbb{I} 
	- i \sin \left( J_1' \alpha_{d \rm D}(\vect{k}) \left(t-\frac{3T}{4}\right) \right) \frac{h_{1,d \rm D} (\vect{k})}{\alpha_{d \rm D}(\vect{k})} \right] \times \non \\
	&& \left[ \cos \left( J_2' \beta_{d \rm D}(\vect{k}) \frac{T}{2} \right) \mathbb{I} 
	- i \sin \left( J_2' \beta_{d \rm D}(\vect{k}) \frac{T}{2} \right) \frac{h_{2,d \rm D} (\vect{k})}{\beta_{d \rm D}(\vect{k})} \right]  
	\left[ \cos \left( J_1' \alpha_{d \rm D}(\vect{k}) \frac{T}{4} \right) \mathbb{I} 
	- i \sin \left( J_1' \alpha_{d \rm D}(\vect{k}) \frac{T}{4} \right) \frac{h_{1,d \rm D} (\vect{k})}{\alpha_{d \rm D}(\vect{k})} \right]  \ .\non \\
	\end{eqnarray}
	After full time-period $T$, we obtain the Floquet operator as
	\begin{eqnarray}
	U_{d \rm D}(\vect{k},T)&=&\left[ \cos \left( J_1' \alpha_{d \rm D}(\vect{k}) \frac{T}{4} \right) \mathbb{I} 
	- i \sin \left( J_1' \alpha_{d \rm D}(\vect{k}) \frac{T}{4} \right) \frac{h_{1,d \rm D} (\vect{k})}{\alpha_{d \rm D}(\vect{k})} \right] \left[ \cos \left( J_2' \beta_{d \rm D}(\vect{k}) \frac{T}{2} \right) \mathbb{I} 
	- i \sin \left( J_2' \beta_{d \rm D}(\vect{k}) \frac{T}{2} \right) \frac{h_{2,d \rm D} (\vect{k})}{\beta_{d \rm D}(\vect{k})} \right]  \non \\
	&& \times \left[ \cos \left( J_1' \alpha_{d \rm D}(\vect{k}) \frac{T}{4} \right) \mathbb{I} 
	- i \sin \left( J_1' \alpha_{d \rm D}(\vect{k}) \frac{T}{4} \right) \frac{h_{1,d \rm D} (\vect{k})}{\alpha_{d \rm D}(\vect{k})} \right] \ .
	\end{eqnarray}
	We can recast $U_{d \rm D}(\vect{k},T)$ in a form such that $U_{d \rm D}(\vect{k},T)=f_{d \rm D}(\vect{k}) \mathbb{I}  -i g_{d \rm D}(\vect{k})$, where we have defined 
	\begin{eqnarray}
	f_{d \rm D}(\vect{k})&=& \cos \left( \frac{\alpha_{d \rm D}(\vect{k}) J_1 }{2} \right)  
	\cos \left( \frac{\beta_{d \rm D}(\vect{k}) J_2 }{2} \right) - 
	\sin \left( \frac{\alpha_{d \rm D}(\vect{k}) J_1 }{2} \right)  
	\sin \left( \frac{\beta_{d \rm D}(\vect{k}) J_2 }{2} \right) 
	\frac{\epsilon_{d \rm D}(\vect{k})}{\alpha_{d \rm D}(\vect{k})\beta_{d \rm D}(\vect{k})} \ , \\
	g_{d \rm D}(\vect{k})&=& 
	\cos^2 \left( \frac{\alpha_{d \rm D}(\vect{k}) J_1 }{2} \right)
	\sin \left( \frac{\beta_{d \rm D}(\vect{k}) J_2 }{2} \right)
	\frac{h_{2,d \rm D} (\vect{k})}{\beta_{d \rm D}(\vect{k})}
	+ \sin \left( \frac{\alpha_{d \rm D}(\vect{k}) J_1 }{2} \right)
	\cos \left( \frac{\beta_{d \rm D}(\vect{k}) J_2 }{2} \right)
	\frac{h_{1,d \rm D} (\vect{k})}{\alpha_{d \rm D}(\vect{k})} \non \\
	&&+
	\sin^2 \left( \frac{\alpha_{d \rm D}(\vect{k}) J_1 }{2} \right)
	\sin \left( \frac{\beta_{d \rm D}(\vect{k}) J_2 }{2} \right) 
	\frac{\left(2 \epsilon_{d \rm D}(\vect{k})-h_{2,d \rm D} (\vect{k})\right)}{\alpha^2_{d \rm  D}(\vect{k})\beta_{d \rm D}(\vect{k})} \ .
	\end{eqnarray}
	The eigenvalue equation for $U_{d \rm D}(\vect{k},T)$ reads: $U_{d \rm D}(\vect{k},T) \ket{\Psi}=\exp[-i E(\vect{k})T] \ket{\Psi}$, which gives us 
	\begin{eqnarray}
	\cos E(\vect{k}) = f_{d \rm D}(\vect{k}) \ ,
	\end{eqnarray}
	with each band being $\frac{N}{2}$-fold degenerate. In our model we have considered eight-band model both in 2D and 3D. Hence, the bands are four-fold degenerate. We denote these bands as 
	$\ket{\Psi_{\pm E_i(\vect{k})}}$, with $+(-)E_i(\vect{k})$ representing unfilled~(filled) bands.
	
	\section{Periodized evolution operators} \label{Sec:S2}
	The time evolution operator in a time periodic system can be decomposed into two parts $U_{d \rm D}(\vect{k},t)=U_{d \rm D,\epsilon}(\vect{k},t) \left[ U_{d \rm D}(\vect{k},T) \right]^{t/T}_\epsilon$. Here, $U_{d \rm D, \epsilon}(\vect{k},t)=U_{d \rm D,\epsilon}(\vect{k},t+T)$ represents the \textit{anomalous} periodized evolution operator encaptulating the dynamics of the system and $\left[ U_{d \rm D}(\vect{k},T) \right]^{t/T}_\epsilon$ represents the \textit{normal} static accumulative part. We can construct $\left[ U_{d \rm D}(\vect{k},T) \right]^{-t/T}_\epsilon$ as follows
	\begin{eqnarray}
	\left[ U_{d \rm D}(\vect{k},T) \right]^{-t/T}_{\epsilon=0}&=& \sum_{i=1}^{N/2} e^{-i (2 \pi - E_i(\vect{k})) t/T}  
	\ket{\Psi_{- E_i(\vect{k})}} 
	\bra{\Psi_{- E_i(\vect{k})}}  
	+ \sum_{i=N/2+1}^{N} e^{-i E_i(\vect{k}) t/T}  \ket{\Psi_{+ E_i(\vect{k})}} 
	\bra{\Psi_{+ E_i(\vect{k})}}\ ,  \\
	\left[ U_{d \rm D}(\vect{k},T) \right]^{-t/T}_{\epsilon=\pi}&=& \sum_{i=1}^{N/2} e^{i E_i(\vect{k}) t/T}  \ket{\Psi_{- E_i(\vect{k})}} 
	\bra{\Psi_{- E_i(\vect{k})}}  + \sum_{i=N/2+1}^{N} e^{-i E_i(\vect{k}) t/T}  \ket{\Psi_{+ E_i(\vect{k})}} 
	\bra{\Psi_{+ E_i(\vect{k})}}  \ .
	\end{eqnarray}
	With the $\left[ U_{d \rm D}(\vect{k},T) \right]^{-t/T}_\epsilon$ in hand, one can obatin the periodized evolution operator using Eqs.~(\ref{U1}), (\ref{U2}), and (\ref{U3}) as
	\begin{equation}
	U_{d \rm D,\epsilon}(\vect{k},t)=U_{d \rm D}(\vect{k},t) \left[ U_{d \rm D}(\vect{k},T) \right]^{-t/T}_\epsilon\ .
	\end{equation}
	
	\section{Dynamical multipole moments}\label{Sec:S3}
	The absence of band physics for $U_{d \rm D, \epsilon}(\vect{k},t)$ enforces us to endeavour for a new quantity to encaptulate the evolution of polarization and other higher moments viz quadrupole, octupole etc, in the interval $t \in \left[0,t\right]$. The dynamical polarization, introduced in Ref.~\cite{Huang2020}, accounts for a comparison of relative motion of a particle between two time intervals and can be defined as 
	\begin{equation}
	\hat{\bar{x}}(t)=\frac{\hat{x}(t)+\hat{x}(0)}{2} \ .
	\end{equation}
	Following Resta's definition~\cite{RestaPRL1998} for position operator $\hat{x}(0)=\hat{x}$, along $x$-direction, for a system obeying periodic boundary condition (PBC) can be written as
	\begin{equation}
	\hat{x}=\sum_{im} \hat{c}_{im}^\dagger \ket{0} e^{-i \Delta_x x_i} \bra{0} \hat{c}_{im} \ ,
	\end{equation}
	where, $\hat{c}_{im}~(\hat{c}_{im}^\dagger)$ represent quasiparticle annihilation~(creation) operator at site $i$ for $m^{\rm th}$ degrees of freedom. We can use the Fourier transformed electronic 
	operator as 
	\begin{equation}
	\hat{c}_{\vect{k}m}=\frac{1}{\sqrt{N}} \sum_{\vect{r}_i} e^{i \vect{k} \cdot \vect{r}_i} \  \hat{c}_{im}   \hspace{4mm} \textrm{and} \hspace{4mm} \hat{c}_{\vect{k}m}^\dagger=\frac{1}{\sqrt{N}} \sum_{\vect{r}_i} e^{i \vect{k} \cdot \vect{r}_i} \  \hat{c}_{im}^\dagger \ ,
	\end{equation}
	with $N=L_x L_y L_z \cdots $, $\vect{r}_i=x_i \vect{e}_x+y_i \vect{e}_y+z_i \vect{e}_z + \cdots$, and $\vect{\Delta}=\frac{2 \pi}{L_x} \vect{e}_x+\frac{2 \pi}{L_y} \vect{e}_y + \frac{2 \pi}{L_z} \vect{e}_z+ \cdots$. We can write $U_{d \rm D, \epsilon}(\vect{k},t)$ as
	\begin{equation}
	U_{d \rm D, \epsilon}(\vect{k},t)=\sum_{\vect{k}mn} \hat{c}_{\vect{k}m}^\dagger \ket{0} \left[U_{d \rm D, \epsilon}(\vect{k},t)\right]_{mn} \bra{0} \hat{c}_{\vect{k}n} \ .
	\end{equation}
	Then one can write $\hat{x}(t)$ in the form
	\begin{eqnarray}
	\hat{x}(t)&=& 
	U^\dagger_{d \rm D, \epsilon}(\vect{k},t) \ 
	\hat{x} \ 
	U_{d \rm D, \epsilon}(\vect{k},t) \non \\
	&=& \sum_{\vect{k}mn} 
	\hat{c}_{\vect{k}+\Delta_xe_x, m}^\dagger \ket{0} \left[U^\dagger_{d \rm D, \epsilon}(\vect{k}+\Delta_x\vect{e}_x,t) 
	U_{d \rm D, \epsilon}(\vect{k},t)\right]_{mn} \bra{0} \hat{c}_{\vect{k}n} \ .
	\end{eqnarray}
	Thus, we obtain average polarization $\hat{\bar{x}}$ as
	\begin{equation}
	\hat{\bar{x}}=\sum_{\vect{k}mn} \hat{c}_{\vect{k}+\Delta_xe_x, m}^\dagger \ket{0} \left[Q_{x,\epsilon,\vect{k}}(t)\right]_{mn} \bra{0} \hat{c}_{\vect{k}n} \ ,
	\end{equation}
	where, we have defined 
	\begin{equation}
	Q_{x,\epsilon,\vect{k}}(t)=\frac{\mathbb{I} + U^\dagger_{d \rm D, \epsilon}(\vect{k}+\Delta_x \vect{e}_x,t) U_{d \rm D, \epsilon}(\vect{k},t)}{2} \ .
	\end{equation}
	The eigen-problem for $\hat{\bar{x}}$ can be solved by considering a $L_{x}^{\rm th}$ power of the same, such that \footnote{Here, $Q_{x,\epsilon,\vect{k}}(t)$'s are not unitary for a finite $L_x$. To perform the numerical calculations, one can do the singular-value decomposition~(SVD), such that $Q_{x,\epsilon,\vect{k}}(t)=UDV^\dagger$ and then redefine $Q_{x,\epsilon,\vect{k}}(t)=UV^\dagger$~\cite{benalcazarprb2017}.}
	\begin{equation}
	\left(\hat{\bar{x}}(t)\right)^{L_x}=\sum_{\vect{k}mn} \hat{c}_{\vect{k}m}^\dagger \ket{0} \left[W_{x,\epsilon,\vect{k}}(t)\right]_{mn} \bra{0} \hat{c}_{\vect{k}n} \ , 
	\end{equation}
	where, we have defined the time-dependent Wilson loop operator as 
	\begin{equation}
	W_{x,\epsilon,\vect{k}}(t)= Q_{x,\epsilon,\vect{k}+(L_x-1)\Delta_x \vect{e}_x}(t) \cdots Q_{x,\epsilon,\vect{k}+\Delta_x \vect{e}_x}(t) Q_{x,\epsilon,\vect{k}}(t) \ .
	\end{equation}
	We can write down the eigenvalue equation for $W_{x,\epsilon,\vect{k}}(t)$ as 
	\begin{equation}
	W_{x,\epsilon,\vect{k}}(t) \ket{\nu_{x,\epsilon , \mu}(\vect{k},t)} = e^{-2 \pi i \nu_{x,\epsilon, \mu}(k_{j \neq x},t)}\ket{\nu_{x,\epsilon , \mu}(\vect{k},t)} \ ,
	\end{equation}
	here, $\mu$ denotes all the $N$ pseudospin degrees of freedom of the Hamiltonian, constituting differnt branches and number of first-order branch equals the number of pseudospin degrees of freedom. The eigenstates $\ket{\nu_{x,\epsilon , \mu}(\vect{k},t)}$ follows the relation $\braket{\nu_{x,\epsilon , \mu_1}(\vect{k},t) | \nu_{x,\epsilon , \mu_2}(\vect{k},t)}=\delta_{\mu_1 \mu_2}$\footnote{Although, the eigenvalues $\nu_{x,\epsilon, \mu}(k_{j \neq x},t)$ are independent irrespective of the choice of the base point $k_x$, but the eigenstates $\ket{\nu_{x,\epsilon , \mu}(\vect{k},t)}$ do depend on the choice 
		of the base point.}. For each branch $\mu$, we can find the eigenvalues of $\hat{\bar{x}}(t)$ by taking a $L_{x}^{\rm th}$ root of $e^{-2 \pi i \nu_{x,\epsilon, \mu}(k_{j \neq x},t)}$ as~\footnote{For numerical stability, one might consider taking a logarithm of the Wilson loop operator to obtain the Wilson Hamiltonian as $H_W=\frac{i}{2 \pi}\log \left[W_{x,\epsilon,\vect{k}}(t)\right]$ and then calculate the eigenvalues and eigenvectors of $H_W$, which coincides with that of the Wilson loop.}
	\begin{equation}
	\hat{\bar{x}}(t) \ket{\psi_{x,\epsilon,\mu}(x_i, k_{j \neq x},t)} = e^{-i \Delta_x \left(x_i + \nu_{x,\epsilon, \mu}(k_{j \neq x},t) \right)} \ket{\psi_{x,\epsilon,\mu}(x_i, k_{j \neq x},t)} \ .
	\end{equation}
	
	The significance of $\nu_{x,\epsilon, \mu}(k_{j \neq x},t)$ can be inferred in way that it bespeaks a relative motion of a particle by $2 \nu_{x,\epsilon, \mu}(k_{j \neq x},t)$ with respect to $x_i$ in the $x$-direction from time $t=0$ to $t$. In contrast to the static counterpart, $\hat{\bar{x}}(t)$ portrays the interference of polarization at two different time-instant $t=0$ and $t$~\cite{Huang2020,benalcazarprb2017}. The following interference pattern $\ket{\psi_{x,\epsilon,\mu}(x_i, k_{j \neq x},t)}$ is given as
	\begin{equation}
	\ket{\psi_{x,\epsilon,\mu}(x_i, k_{j \neq x},t)}= \frac{1}{\sqrt{L_x}} \sum_{k_x m} \hat{c}^\dagger_{\vect{k}m} \ket{0}  e^{i k_x x_i} \left[ \nu_{x,\epsilon , \mu} (\vect{k},t) \right]_m \ .  
	\end{equation}
	\begin{figure}[]
		\centering
		\subfigure{\includegraphics[width=0.75\textwidth]{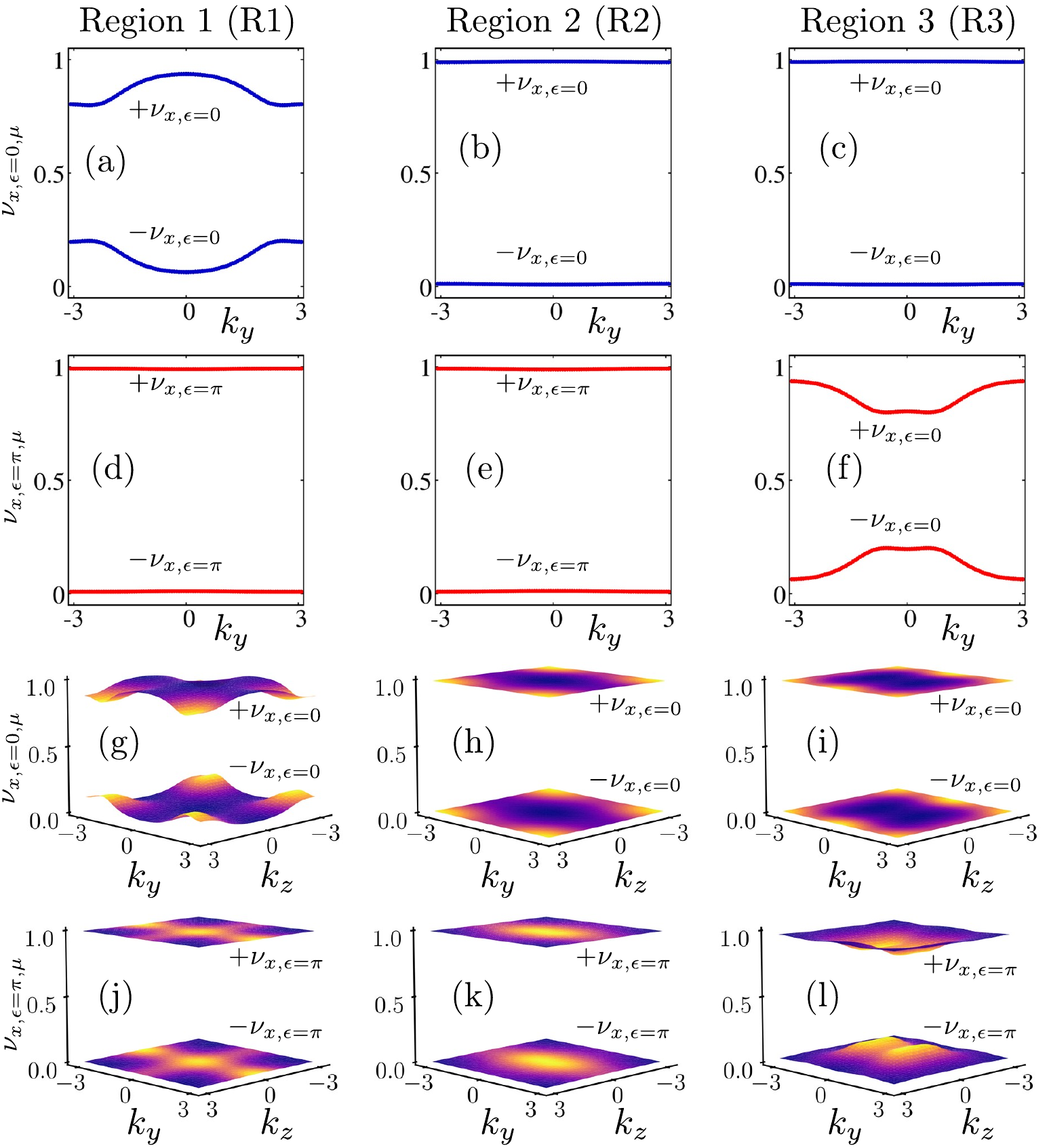}}
		\caption{We depict the dynamical first-order polarization branch for 2D FSOTSC while the system is in R1, R2, and R3 (see maintext), in panels (a), (b), and (c) respectively for $0$-gap and panels 
			(d), (e), and (f), respectively for $\pi$-gap. We repeat the same for 3D system and show the dynamical first-order polarization as a function of $k_y$ and $k_z$, keeping the system in R1, R2, and R3 
			in panels (g), (h), and (i), respectively for $0$-gap and panels (j), (k), and (l), respectively for $\pi$-gap.
		}
		\label{firstorderpol}
	\end{figure}
	\subsection{Dynamical quadrupolar motion}
	To proceed, out of $N$ branches of $\nu_{x,\epsilon, \mu}(k_{j \neq x},t)$, we can group them into two separable sets $\pm \nu_{x,\epsilon}$ and the second-order polarization corresponds to the motion of the particle perpendicular to $x$-direction within each branch set $\pm\nu_{x,\epsilon}$. We choose $y$-direction to be the perpendicular direction and write down the mean polarization along $y$-direction as 
	\begin{eqnarray}
	\hat{\bar{y}}&=& \frac{\hat{x}(t)+\hat{y}(0)}{2} \non \\
	&=& \sum_{\vect{k}mn} \hat{c}_{\vect{k}+\Delta_ye_y, m}^\dagger \ket{0} \left[Q_{y,\epsilon,\vect{k}}(t)\right]_{mn} \bra{0} \hat{c}_{\vect{k}n}
	\end{eqnarray}
	where, $Q_{y,\epsilon,\vect{k}}(t)$ is given as
	\begin{equation}
	Q_{y,\epsilon,\vect{k}}(t)=\frac{\mathbb{I} + U^\dagger_{d \rm D,\epsilon}(\vect{k}+\Delta_y\vect{e}_y,t) U_{d \rm D,\epsilon}(\vect{k},t)}{2} \ .
	\end{equation}
	
	We define the branch projector operator as 
	\begin{eqnarray}
	P_{\pm \nu_{x, \epsilon}} (t)&=& \sum_{x_i, k_{j \neq x}, \mu \in \pm \nu_{x, \epsilon}} \ket{\psi_{x,\epsilon,\mu}(x_i, k_{j \neq x},t)} \bra{\psi_{x,\epsilon,\mu}(x_i, k_{j \neq x},t)} \non \\
	&=& \sum_{\vect{k}mn, \mu \in \pm \nu_{x, \epsilon}} \hat{c}^\dagger_{\vect{k}m} \ket{0} \left[ \nu_{x,\epsilon , \mu}(\vect{k},t) \right]_m \left[ \nu_{x,\epsilon , \mu} (\vect{k},t)\right]_n^* \bra{0} \hat{c}_{\vect{k}n} \ .
	\end{eqnarray}
	We introduce the dynamical branch creation and annihilation operator as 
	\begin{eqnarray}
	\hat{\gamma}_{\vect{k} \epsilon \mu } (t) &=& \sum_m \hat{c}_{\vect{k}m} \left[\nu_{x,\epsilon , \mu} (\vect{k},t) \right]_m^* \non \\
	\hat{\gamma}^\dagger_{\vect{k} \epsilon \mu } (t) &=& \sum_m \hat{c}^\dagger_{\vect{k}m} \left[\nu_{x,\epsilon , \mu} (\vect{k},t) \right]_m \ ,
	\end{eqnarray}
	with $\left\{\hat{\gamma}_{\vect{k} \epsilon \mu_1 } (t),\hat{\gamma}_{\vect{k'} \epsilon \mu_2 }^\dagger (t)\right\}=\delta_{\vect{k}\vect{k'}}\delta_{\mu_1 \mu_2}$. Thus we can rewrite the projector as 
	\begin{equation}
	P_{\pm \nu_{x, \epsilon}} (t)= \sum_{\vect{k},\mu \in \pm \nu_{x, \epsilon}} \hat{\gamma}^\dagger_{\vect{k} \epsilon \mu } (t) \ket{0} \bra{0} \hat{\gamma}_{\vect{k} \epsilon \mu } (t) \ .
	\end{equation}
	
	Afterwards, we project $\hat{\bar{y}}$ to the branch set $\pm \nu_{x,\epsilon}$ to obtain the dynamical second-order polarization as 
	\begin{eqnarray}
	\hat{\bar{y}}^{\pm \nu_{x,\epsilon}} (t)&=& P_{\pm \nu_{x, \epsilon}} (t) \ \hat{\bar{y}} (t) \ P_{\pm \nu_{x, \epsilon}} (t) \non \\
	&=& \sum_{\vect{k},\mu_1,\mu_2 \in \pm \nu_{x, \epsilon}} \hat{\gamma}^\dagger_{\vect{k} + \Delta_y \vect{e}_y , \epsilon \mu_1 } (t) \ket{0} \left[Q_{y,\epsilon,\vect{k}}^{\pm \nu_{x, \epsilon}} (t)\right]_{\mu_1 \mu_2} \bra{0}  \hat{\gamma}_{\vect{k} \epsilon \mu_2 } (t)
	\end{eqnarray}
	where, we have defined $Q_{y,\epsilon,\vect{k}}^{\pm \nu_{x, \epsilon}} (t)$ as
	\begin{equation}
	\left[Q_{y,\epsilon,\vect{k}}^{\pm \nu_{x, \epsilon}} (t)\right]_{\mu_1 \mu_2} = \sum_{m n} \left[\nu_{x,\epsilon , \mu_1} (\vect{k}+\Delta_y \vect{e}_y,t) \right]^*_m  \left[Q_{y,\epsilon,\vect{k}}(t)\right]_{mn} \left[\nu_{x,\epsilon , \mu_2} (\vect{k},t) \right]_n \ .
	\end{equation}
	
	The dynamical second-order polarization (dynamical quadrupole) problem can be solved by considering $L_{y}^{\rm th}$ power of $\hat{\bar{y}}^{\pm \nu_{x,\epsilon}} (t)$, such that
	\begin{eqnarray}
	\left(\hat{\bar{y}}^{\pm \nu_{x,\epsilon}} (t)\right)^{L_y}=\sum_{\vect{k},\mu_1,\mu_2 \in \pm \nu_{x, \epsilon}} \gamma_{\vect{k} \epsilon \mu_1 }^\dagger(t) \ket{0} \left[W_{y,\epsilon,\vect{k}}^{\pm \nu_{x,\epsilon}} (t) \right]_{\mu_1\mu_2} \bra{0} \gamma_{\vect{k} \epsilon \mu_2 } (t) \ ,
	\end{eqnarray}
	where, the time dependent first-order nested Wilson loop $W_{y,\epsilon,\vect{k}}^{\pm \nu_{x,\epsilon}} (t)$ is given as
	\begin{equation}
	W_{y,\epsilon,\vect{k}}^{\pm \nu_{x,\epsilon}} (t)= Q_{y,\epsilon,\vect{k}+(L_y-1)\Delta_y \vect{e}_y}^{\pm \nu_{x,\epsilon}}(t) \cdots Q_{y,\epsilon,\vect{k}+\Delta_y \vect{e}_y}^{\pm \nu_{x,\epsilon}}(t) Q_{y,\epsilon,\vect{k}}^{\pm \nu_{x,\epsilon}}(t) \ .
	\end{equation}
	
	One can procure differnt quadrupole branches $\mu'$, by diagonalizing $W_{y,\epsilon,\vect{k}}^{\pm \nu_{x,\epsilon}} (t)$ as 
	\begin{eqnarray}
	W_{y,\epsilon,\vect{k}}^{\pm \nu_{x,\epsilon}} (t) \ket{\nu_{y,\epsilon , \mu'}^{\pm \nu_{x,\epsilon}}(\vect{k},t)} = e^{-2 \pi i \nu_{y,\epsilon, \mu'}^{\pm \nu_{x,\epsilon}}(k_{j \neq y},t)}\ket{\nu_{y,\epsilon , \mu'}^{\pm \nu_{x,\epsilon}}(\vect{k},t)} \ .
	\end{eqnarray}
	
	The dynamical quadrupolar eigenproblem can be solved by procuring a $L_{y}^{\rm th}$ root of $e^{-2 \pi i \nu_{y,\epsilon, \mu'}^{\pm \nu_{x,\epsilon}}(k_{j \neq y},t)}$ as
	\begin{equation}
	\hat{\bar{y}}^{\pm \nu_{x,\epsilon}} (t) \ket{\chi_{y,\epsilon,\mu'}^{\pm \nu_{x,\epsilon}}(y_i, k_{j \neq y},t)}= e^{-i \Delta_y \left(y_i + \nu_{y,\epsilon, \mu'}^{\pm \nu_{x,\epsilon}}(k_{j \neq y},t) \right)} \ket{\chi_{y,\epsilon,\mu'}^{\pm \nu_{x,\epsilon}}(y_i, k_{j \neq y},t)} \ ,
	\end{equation}
	with the dynamical quadrupolar interference pattern is given by
	\begin{equation}
	\ket{\chi_{y,\epsilon,\mu'}^{\pm \nu_{x,\epsilon}}(y_i, k_{j \neq y},t)}= \frac{1}{\sqrt{L_y}} \sum_{k_y , \mu \in \pm \nu_{x, \epsilon}} \hat{\gamma}^\dagger_{\vect{k} \mu} \ket{0}  e^{i k_y y_i} \left[ \nu_{y,\epsilon , \mu'}^{\pm \nu_{x,\epsilon}}(\vect{k},t) \right]_\mu \ .  
	\end{equation}
	
	Limited to a 2D system, one can obtain the average quadrupolar motion as 
	\begin{equation}
	\boxed{\langle \nu_{y,\epsilon , \mu'}^{\pm \nu_{x,\epsilon}} \rangle (t)=\frac{1}{L_x} \sum_{k_x} \nu_{y,\epsilon , \mu'}^{\pm \nu_{x,\epsilon}} (k_x,t)} \ .
	\end{equation}
	\begin{figure}[]
		\centering
		\subfigure{\includegraphics[width=0.98\textwidth]{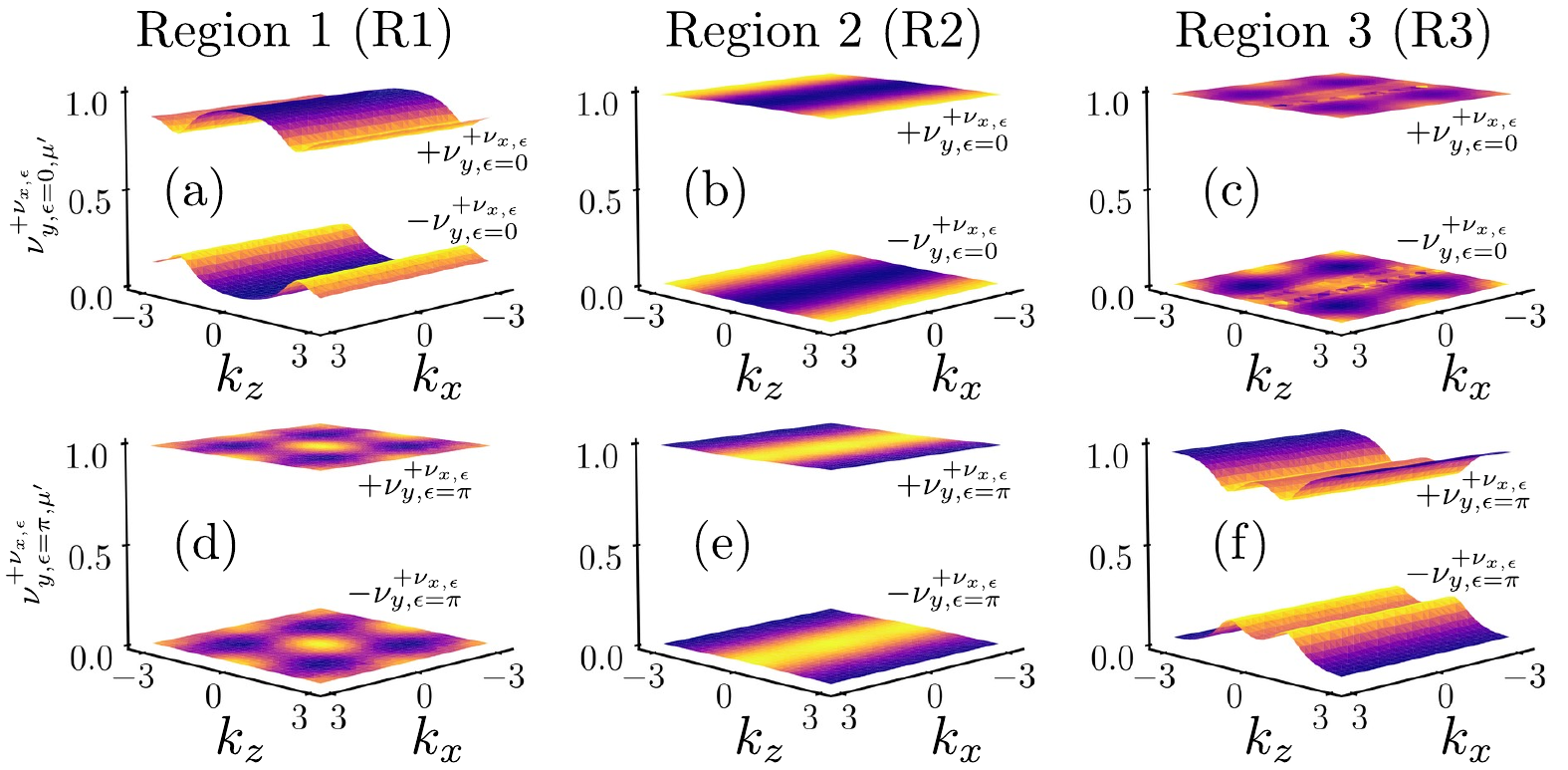}}
		\caption{We demonstrate the dynamical second-order polarization branch for 3D FTOTSC while the system is in R1, R2, and R3. Here, panels (a), (b), and (c), respectively correspond to $0$-gap 
			and (d), (e), and (f), respectively refer to the $\pi$-gap.
		}
		\label{secondorderpol}
	\end{figure}
	\subsection{Dynamical octupolar motion}
	We have obtained $\frac{N}{2}$ numbers of $\nu_{y,\epsilon, \mu'}^{\pm\nu_{x,\epsilon}}(k_{j \neq y},t)$ branches denoted by $\mu'$ within each branch $\pm\nu_{x,\epsilon}$, out of which, we construct two groups by identifying them as $\pm \nu_{y,\epsilon}^{\pm\nu_{x,\epsilon}}$. We define the quadrupolar branch projector as 
	\begin{eqnarray}
	P_{\pm \nu_{y,\epsilon }^{\pm \nu_{x,\epsilon}}} (t)&=& \sum_{\substack{y_i, k_{j \neq y}, \\ \mu' \in \pm \nu_{y,\epsilon }^{\pm \nu_{x,\epsilon}}}} \ket{\chi_{y,\epsilon,\mu'}^{\pm \nu_{x,\epsilon}}(y_i, k_{j \neq y},t)}
	\bra{\chi_{y,\epsilon,\mu'}^{\pm \nu_{x,\epsilon}}(y_i, k_{j \neq y},t)} \non \\
	&=& \sum_{\substack{\vect{k}mn\mu_1\mu_2, \\ \mu' \in \pm \nu_{y,\epsilon }^{\pm \nu_{x,\epsilon}}}} \hat{c}^\dagger_{\vect{k}m} \ket{0} 
	\left[ \nu_{x,\epsilon , \mu_1}(\vect{k},t) \right]_m 
	\left[ \nu_{y,\epsilon , \mu'}^{\pm \nu_{x,\epsilon}}(\vect{k},t) \right]_{\mu_1}
	\left[ \nu_{x,\epsilon , \mu_2}(\vect{k},t) \right]_n^* 
	\left[ \nu_{y,\epsilon , \mu'}^{\pm \nu_{x,\epsilon}}(\vect{k},t) \right]_{\mu_2}^*
	\bra{0} \hat{c}_{\vect{k}n} \ .
	\end{eqnarray}
	
	We introduce the second-order dynamical branch creation and  annihilation operator as 
	\begin{eqnarray}
	\hat{\eta}_{\vect{k}\epsilon \mu'}&=&\sum_{m\mu} 
	\hat{c}_{\vect{k}m} 
	\left[ \nu_{x,\epsilon , \mu}(\vect{k},t) \right]_m^* 
	\left[ \nu_{y,\epsilon , \mu'}^{\pm \nu_{x,\epsilon}}(\vect{k},t) \right]_{\mu}^* \ , \non \\
	\hat{\eta}_{\vect{k}\epsilon\mu'}^\dagger&=&\sum_{m\mu} 
	\hat{c}^\dagger_{\vect{k}m} 
	\left[ \nu_{x,\epsilon , \mu}(\vect{k},t) \right]_m 
	\left[ \nu_{y,\epsilon , \mu'}^{\pm \nu_{x,\epsilon}}(\vect{k},t) \right]_{\mu}\ ,
	\end{eqnarray}
	with $\left\{\hat{\eta}_{\vect{k} \epsilon \mu'_1 } (t),\hat{\eta}_{\vect{k'} \epsilon \mu'_2 }^\dagger (t)\right\}=\delta_{\vect{k}\vect{k'}}\delta_{\mu'_1 \mu'_2}$. We can rewrite the projector as 
	\begin{equation}
	P_{\pm \nu_{y,\epsilon }^{\pm \nu_{x,\epsilon}}} (t)= \sum_{\vect{k},\mu' \in \pm \nu_{y,\epsilon }^{\pm \nu_{x,\epsilon}}} \hat{\eta}^\dagger_{\vect{k} \epsilon \mu' } (t) \ket{0} \bra{0} \hat{\eta}_{\vect{k} \epsilon \mu' } (t) \ .
	\end{equation}
	
	The dynamical third-order polarization can be extracted by considering the motion of the particle perpendicular to both $x$ and $y$-directions (\ie along $z$-direction), projected to the branch set $\pm \nu_{y,\epsilon }^{\pm \nu_{x,\epsilon}}$. The mean polarization along $z$-direction is given as 
	\begin{eqnarray}
	\hat{\bar{z}}&=& \frac{\hat{z}(t)+\hat{z}(0)}{2} \non \\
	&=& \sum_{\vect{k}mn} \hat{c}_{\vect{k}+\Delta_ze_z, m}^\dagger \ket{0} \left[Q_{z,\epsilon,\vect{k}}(t)\right]_{mn} \bra{0} \hat{c}_{\vect{k}n}\ ,
	\end{eqnarray}
	where, $Q_{z,\epsilon,\vect{k}}(t)$ is given as
	\begin{equation}
	Q_{z,\epsilon,\vect{k}}(t)=\frac{\mathbb{I} + U^\dagger_\epsilon(\vect{k}+\Delta_z \vect{e}_z,t) U_\epsilon(\vect{k},t)}{2} \ .
	\end{equation}
	
	Subsequently, we project $\hat{\bar{z}}$ to the branch set $\pm \nu_{y,\epsilon }^{\pm \nu_{x,\epsilon}}$ to obtain the dynamical third-order polarization as 
	\begin{eqnarray}
	\hat{\bar{z}}^{\pm \nu_{y,\epsilon }^{\pm \nu_{x,\epsilon}}} (t)&=& P_{\pm \nu_{y,\epsilon }^{\pm \nu_{x,\epsilon}}} (t) \ \hat{\bar{z}} (t) \ P_{\pm \nu_{y,\epsilon }^{\pm \nu_{x,\epsilon}}} (t) \non \\
	&=& \sum_{\substack{\vect{k},  \mu'_1,\mu'_2 \in \pm \nu_{y,\epsilon }^{\pm \nu_{x,\epsilon}}}} \hat{\eta}^\dagger_{\vect{k}+ \Delta_z \vect{e}_z, \epsilon \mu'_1 } (t) \ket{0} \left[Q_{z,\epsilon,\vect{k}}^{\pm \nu_{y,\epsilon }^{\pm \nu_{x,\epsilon}}} (t)\right]_{\mu'_1 \mu'_2} \bra{0}  \hat{\eta}_{\vect{k} \epsilon \mu'_2 } (t)\ ,
	\end{eqnarray}
	where, we have defined $Q_{z,\epsilon,\vect{k}}^{\pm \nu_{y,\epsilon }^{\pm \nu_{x,\epsilon}}} (t)$ as
	\begin{equation}
	\left[Q_{z,\epsilon,\vect{k}}^{\pm \nu_{y,\epsilon }^{\pm \nu_{x,\epsilon}}} (t)\right]_{\mu'_1 \mu'_2} =
	\sum_{\substack{m n \mu_1 \mu_2 }} 
	\left[ \nu_{y,\epsilon , \mu'_1}^{\pm \nu_{x,\epsilon}}(\vect{k}+\Delta_z \vect{e}_z,t) \right]_{\mu_1}^*
	\left[ \nu_{x,\epsilon , \mu_1}(\vect{k}+\Delta_z \vect{e}_z,t) \right]_m^* 
	\left[Q_{z,\epsilon,\vect{k}}(t)\right]_{mn}
	\left[ \nu_{x,\epsilon , \mu_2}(\vect{k},t) \right]_n
	\left[ \nu_{y,\epsilon , \mu'_2}^{\pm \nu_{x,\epsilon}}(\vect{k},t) \right]_{\mu_2}\ .
	\end{equation}
	
	Afterwards, this dynamical third-order polarization (dynamical octupolar) problem can be solved by considering $L_{z}^{\rm th}$ power of $\hat{\bar{z}}^{\pm \nu_{y,\epsilon }^{\pm \nu_{x,\epsilon}}} (t)$, such that
	\begin{eqnarray}
	\left(\hat{\bar{z}}^{\pm \nu_{y,\epsilon }^{\pm \nu_{x,\epsilon}}}(t)\right)^{L_z}=
	\sum_{\vect{k},\mu'_1,\mu'_2 \in \pm \nu_{y,\epsilon }^{\pm \nu_{x,\epsilon}}} 
	\eta_{\vect{k} \epsilon \mu'_1 }^\dagger(t) 
	\ket{0} 
	\left[W_{z,\epsilon,\vect{k}}^{\pm \nu_{y,\epsilon }^{\pm \nu_{x,\epsilon}}} (t)\right]_{\mu'_1 \mu'_2} 
	\bra{0} 
	\eta_{\vect{k} \epsilon \mu'_2 } (t) \ ,
	\end{eqnarray}
	where, the time dependent second-order nested Wilson loop $W_{z,\epsilon,\vect{k}}^{\pm \nu_{y,\epsilon }^{\pm \nu_{x,\epsilon}}} (t)$ is given as
	\begin{equation}
	W_{z,\epsilon,\vect{k}}^{\pm \nu_{y,\epsilon }^{\pm \nu_{x,\epsilon}}} (t)
	= 
	Q_{z,\epsilon,\vect{k}+(L_z-1)\Delta_z \vect{e}_z}^{\pm \nu_{y,\epsilon }^{\pm \nu_{x,\epsilon}}}(t) 
	\cdots 
	Q_{z,\epsilon,\vect{k}+\Delta_z \vect{e}_z}^{\pm \nu_{y,\epsilon }^{\pm \nu_{x,\epsilon}}}(t) 
	Q_{z,\epsilon,\vect{k}}^{\pm \nu_{y,\epsilon }^{\pm \nu_{x,\epsilon}}}(t) \ .
	\end{equation}
	
	Thus, one can extract different octupolar branches $\mu''$, by diagonalizing $W_{z,\epsilon,\vect{k}}^{\pm \nu_{y,\epsilon }^{\pm \nu_{x,\epsilon}}} (t)$ as 
	\begin{eqnarray}
	W_{z,\epsilon,\vect{k}}^{\pm \nu_{y,\epsilon }^{\pm \nu_{x,\epsilon}}} (t)
	\ket{\nu_{z,\epsilon , \mu''}^{\pm \nu_{y,\epsilon }^{\pm \nu_{x,\epsilon}}}(\vect{k},t)} 
	= 
	e^{-2 \pi i \nu_{z,\epsilon, \mu''}^{\pm \nu_{y,\epsilon }^{\pm \nu_{x,\epsilon}}}(k_{j \neq z},t)}
	\ket{\nu_{z,\epsilon , \mu''}^{\pm \nu_{y,\epsilon }^{\pm \nu_{x,\epsilon}}}(\vect{k},t)} \ .
	\end{eqnarray}
	
	Therefore, the dynamical quadrupolar eigenproblem can be solved by taking a $L_{z}^{\rm th}$ root of $e^{-2 \pi i \nu_{z,\epsilon, \mu''}^{\pm \nu_{y,\epsilon }^{\pm \nu_{x,\epsilon}}}(k_{j \neq z},t)}$, 
	such that
	\begin{equation}
	\hat{\bar{z}}^{\pm \nu_{y,\epsilon }^{\pm \nu_{x,\epsilon}}}(t) \ket{\zeta_{z,\epsilon,\mu''}^{\pm \nu_{y,\epsilon }^{\pm \nu_{x,\epsilon}}}(z_i, k_{j \neq z},t)}
	= 
	e^{-2 \pi i \nu_{z,\epsilon, \mu''}^{\pm \nu_{y,\epsilon }^{\pm \nu_{x,\epsilon}}}(k_{j \neq z},t)}
	\ket{\zeta_{z,\epsilon,\mu''}^{\pm \nu_{y,\epsilon }^{\pm \nu_{x,\epsilon}}}(z_i, k_{j \neq z},t)} \ .
	\end{equation}
	with the dynamical octupolar interference pattern is given as 
	\begin{equation}
	\ket{\zeta_{z,\epsilon,\mu''}^{\pm \nu_{y,\epsilon }^{\pm \nu_{x,\epsilon}}}(z_i, k_{j \neq z},t)}
	= \frac{1}{\sqrt{L_z}} \sum_{k_z , \mu' \in \pm \nu_{y,\epsilon }^{\pm \nu_{x,\epsilon}}} \hat{\eta}^\dagger_{\vect{k} \mu'} \ket{0}  e^{i k_z z_i} \left[ \nu_{z,\epsilon , \mu''}^{\pm \nu_{y,\epsilon }^{\pm \nu_{x,\epsilon}}}(\vect{k},t) \right]_{\mu'} \ .  
	\end{equation}
	
	Finally, for a 3D system, one can obtain the average octupolar motion as 
	\begin{equation}
	\boxed {\langle \nu_{z,\epsilon, \mu''}^{\pm \nu_{y,\epsilon }^{\pm \nu_{x,\epsilon}}} \rangle (t)=\frac{1}{L_x L_y} \sum_{k_x k_y} \nu_{z,\epsilon, \mu''}^{\pm \nu_{y,\epsilon }^{\pm \nu_{x,\epsilon}}} (k_x,k_y,t)} \ .
	\end{equation}
	
	\section{Quasi-static multipole moments from the Floquet operator}\label{Sec:S4}
	Here, we briefly discuss the outlines to obtain the quasi-static multipole moments from Floquet operator $U_{d \rm D}(\vect{k},T)$ employing the nested Wilson loop technique~\cite{benalcazarprb2017,Ni2020}. We construct the Wilson loop operator as 
	\begin{equation}
	W_{x,{\rm Flq},\vect{k}}= F_{x,\vect{k}+(L_x-1)\Delta_x \vect{e}_x}(t) \cdots F_{x,\vect{k}+\Delta_x \vect{e}_x} F_{x,\vect{k}} \ ,
	\end{equation}
	where, we have defined $\left[F_{x,\vect{k}}\right]_{mn}=\braket{ \Psi_m (\vect{k}+\Delta_x \vect{e}_x) | \Psi_n (\vect{k})}$ with $\ket{\Psi(\vect{k}) }$'s being the occupied quasi-energy states of the 
	Floquet operator $U_{d \rm D}(\vect{k},T)$. The eigenvalue equation for $W_{x,{\rm Flq},\vect{k}}$ is given as
	\begin{equation}
	W_{x,{\rm Flq},\vect{k}} \ket{\nu_{x, \rm Flq , \mu}(\vect{k})} = e^{-2 \pi i \nu_{x,\rm Flq, \mu}(k_{j \neq x})}\ket{\nu_{x, \rm Flq , \mu}(\vect{k})} \ ,
	\end{equation}
	where, $\nu_{x,\rm Flq, \mu}(k_{j \neq x})$ represents the first-order Wannier sector polarization. The number of branches for $\nu_{x,\rm Flq, \mu}(k_{j \neq x})$ is $\frac{N}{2}$ unlike the dynamical first-order polarization whose number is equal to the total number of degrees (\ie $N$) present in the Hamiltonian. We can divide the first-order polarization in two sectors as $\pm \nu_{x}$. Within each branch 
	$\pm \nu_{x}$, one can construct the first-order nested Wilson loop as~\cite{benalcazarprb2017,Ni2020}
	\begin{equation}
	W_{y,\rm Flq,\vect{k}}^{\pm \nu_{x}}= F_{y,\vect{k}+(L_y-1)\Delta_y \vect{e}_y}^{\pm \nu_{x}} \cdots F_{y,\vect{k}+\Delta_y \vect{e}_y}^{\pm \nu_{x}} F_{y,\vect{k}}^{\pm \nu_{x}} \ ,
	\end{equation}
	where, $F_{y,\vect{k}}^{\pm \nu_{x}}$ is defined as
	\begin{equation}
	\left[F_{y,\vect{k}}^{\pm \nu_{x}} \right]_{\mu_1 \mu_2} = \sum_{m n} \left[\nu_{x, \rm Flq , \mu_1} (\vect{k}+\Delta_y \vect{e}_y) \right]^*_m  \left[F_{y,\vect{k}}\right]_{mn} \left[\nu_{x,\rm Flq , \mu_2} (\vect{k}) \right]_n \ ,
	\end{equation}
	where, we have defined $\left[F_{y,\vect{k}}\right]_{mn}=\braket{ \Psi_m (\vect{k}+\Delta_y \vect{e}_y) | \Psi_n (\vect{k})}$. The eigenvalue equation for $W_{y,\rm Flq,\vect{k}}^{\pm \nu_{x}}$ is given as
	\begin{eqnarray}
	W_{y,\rm Flq,\vect{k}}^{\pm \nu_{x}}  \ket{\nu_{y,\rm Flq , \mu'}^{\pm \nu_{x}}(\vect{k})} = e^{-2 \pi i \nu_{y,\rm Flq, \mu'}^{\pm \nu_{x}}(k_{j \neq y})}\ket{\nu_{y,\rm Flq , \mu'}^{\pm \nu_{x}}(\vect{k})} \ .
	\end{eqnarray}
	
	The number of branch $\nu_{y,\rm Flq, \mu'}^{\pm \nu_{x}}(k_{j \neq y})$ for the quasi-static second-order polarization is half of that of the dynamical second-order polarization. Limited to 2D, the average second-order polarization~(first-order nested polarization) for the $\mu^{\prime \rm th}$ branch is given as
	\begin{equation}
	\langle \nu_{y,\rm Flq, \mu'}^{\pm \nu_{x}}\rangle = \frac{1}{L_x} \sum_{k_x} \nu_{y,\rm Flq, \mu'}^{\pm \nu_{x}}(k_x) \ .
	\end{equation} 
	
	We proceed further and construct the second-order nested Wilson loop in the sector $\pm \nu_{y,\rm Flq, \mu'}^{\pm \nu_{x}}(k_{j \neq y})$ as~\cite{Ni2020}
	\begin{equation}
	W_{z,\rm Flq,\vect{k}}^{\pm \nu_{y }^{\pm \nu_{x}}}
	= 
	F_{z,\vect{k}+(L_z-1)\Delta_z \vect{e}_z}^{\pm \nu_{y}^{\pm \nu_{x}}} 
	\cdots 
	F_{z,\vect{k}+\Delta_z \vect{e}_z}^{\pm \nu_{y}^{\pm \nu_{x}}} 
	F_{z,\vect{k}}^{\pm \nu_{y}^{\pm \nu_{x}}}\ ,
	\end{equation}
	where, we have introduced $F_{z,\vect{k}}^{\pm \nu_{y}^{\pm \nu_{x}}}$ as
	\begin{equation}
	\left[F_{z,\vect{k}}^{\pm \nu_{y }^{\pm \nu_{x}}} \right]_{\mu'_1 \mu'_2} =
	\sum_{\substack{m n \mu_1 \mu_2  }} 
	\left[ \nu_{y,\rm Flq , \mu'_1}^{\pm \nu_{x}}(\vect{k}+\Delta_z\vect{e}_z) \right]_{\mu_1}^*
	\left[ \nu_{x,\rm Flq, \mu_1}(\vect{k}+\Delta_z\vect{e}_z) \right]_m^* 
	\left[F_{z,\vect{k}}\right]_{mn}
	\left[ \nu_{x,\rm Flq , \mu_2}(\vect{k}) \right]_n
	\left[ \nu_{y,\rm Flq , \mu'_2}^{\pm \nu_{x}}(\vect{k}) \right]_{\mu_2} \ ,
	\end{equation}
	here, we define $\left[F_{z,\vect{k}}\right]_{mn}=\braket{ \Psi_m (\vect{k}+\Delta_z \vect{e}_z) | \Psi_n (\vect{k})}$. The eigenvalue equation for $W_{z,\rm Flq,\vect{k}}^{\pm \nu_{y }^{\pm \nu_{x}}}$ 
	is given as
	\begin{equation}
	W_{z,\rm Flq,\vect{k}}^{\pm \nu_{y}^{\pm \nu_{x}}}
	\ket{\nu_{z,\rm Flq , \mu''}^{\pm \nu_{y }^{\pm \nu_{x}}}(\vect{k})} 
	= 
	e^{-2 \pi i \nu_{z,\rm Flq, \mu''}^{\pm \nu_{y}^{\pm \nu_{x}}}(k_{j \neq z})}
	\ket{\nu_{z,\rm Flq , \mu''}^{\pm \nu_{y }^{\pm \nu_{x}}}(\vect{k})}  \ .
	\end{equation}
	
	Therefore, the third-order polarization is characterized by $\nu_{z,\rm Flq, \mu''}^{\pm \nu_{y}^{\pm \nu_{x}}}(k_{j \neq z})$. For a 3D system, we obtain the average third-order polarization for 
	$\mu^{''\rm th}$ branch as
	\begin{equation}
	\langle \nu_{z,\rm Flq, \mu''}^{\pm \nu_{y}^{\pm \nu_{x}}}\rangle = \frac{1}{L_xL_y} \sum_{k_x,k_y} \nu_{z,\rm Flq, \mu''}^{\pm \nu_{y}^{\pm \nu_{x}}}(k_x,k_y) \ .
	\end{equation}
	Note that, this quasi-static multipole moment can only capture the topological character of 0-quasi-energy modes. 
	
	\section{Realization of Floquet second-order topological superconductors in 3D}\label{Sec:S5}
	\begin{figure}[H]
		\centering
		\subfigure{\includegraphics[width=0.98\textwidth]{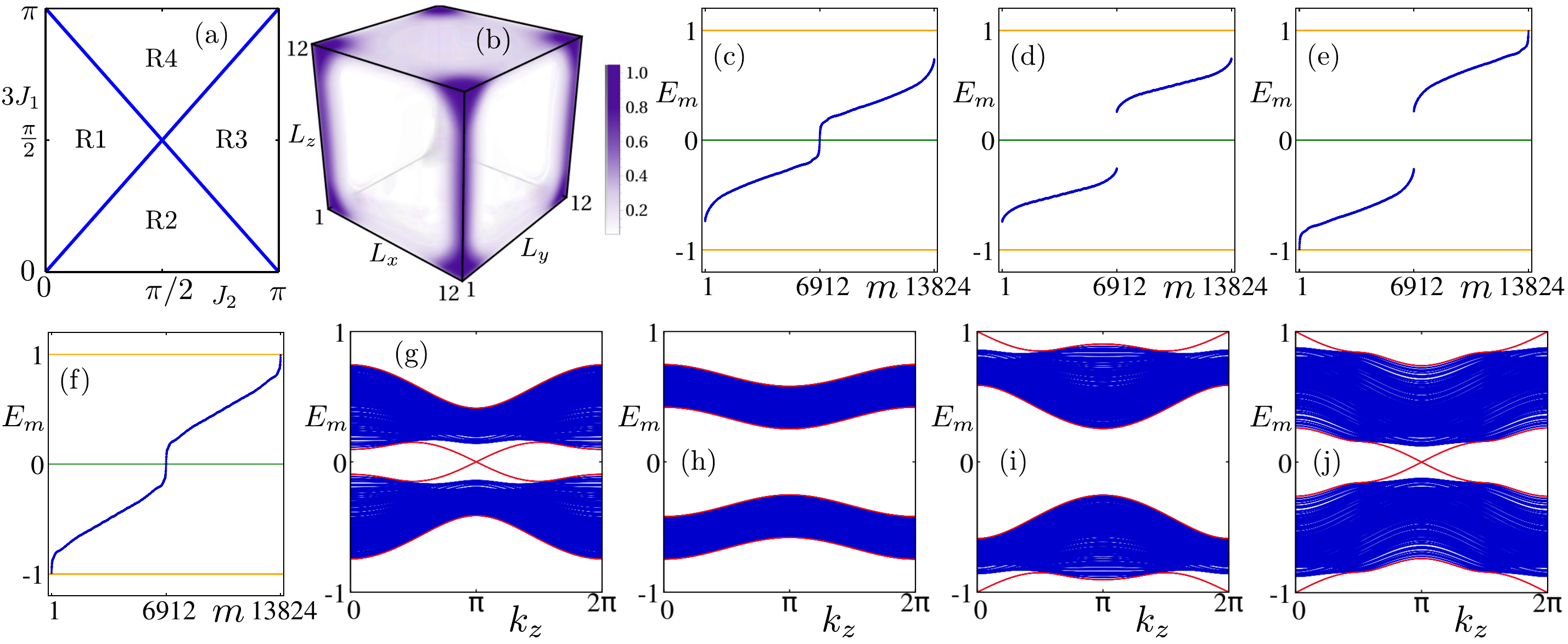}}
		\caption{(a) The phase diagram is depicted in the parameter space $J_1$ and $J_2$ for 3D FSOTSC. (b) The LDOS is demonstrated as a function of the system dimension ($L_{x}\times L_{y}$) 
			for quasienergy $E_m=0, \pm \pi$ while the system is in R4. The quasienergy spectra, considering a finite~(rod) geometry, are shown for R1, R2, R3, and R4 in panels (c), (d), (e), and (f)~((g), (h), (i),  
			and (j)), respectively. See text for discussion. We choose the same parameter set for $(J_1, J_2)$ as used in Fig. 2 of the main text. We set $\Delta_1=1.0$.
		}
		\label{3DSOTSC}
	\end{figure}
	The model introduced in the main text to realize the 3D FTOTSC can also attain the 3D FSOTSC phase. To accomplish the same, we set the amplitude of the $d_{3z^2-r^2}$ pairing to zero (\ie $\Delta_2=0.0$) and retain only the $d_{x^2-y^2}$ pairing. However, the phase diagram remains the same due to the absence of the pairing term in the phase boundary equation (see Eq. (4) of the main text). We show the phase diagram in the $J_1-J_2$ plane in Fig.~\ref{3DSOTSC}~(a). The phase diagram is divided into four parts - region-1~(R1), region-2~(R2), region-3~(R3), and region-4~(R4). The trademark of 3D FTOTSC is the presence of Majorana hinge modes~(MHMs) along the hinges of the system. We depict the footprints of the MHMs in the local density of states~(LDOS) (see Fig.~\ref{3DSOTSC}~(b)) for $E_m=0,\pm \pi$, while the system is in R4. The quasienergy spectra $E_m$, considering open boundary condition (OBC) along all three directions, are shown in Figs.~\ref{3DSOTSC}~(c), (d), (e), and (f) while the system is in R1, R2, R3, and R4, respectively. To obtain the information regarding the dispersive nature of the MHMs, we resort to rod geometry \ie OBC 
	along two directions~($x$ and $y$-direction) and periodic boundary condition (PBC) along the remaining direction~($z$-direction) and show the corresponding quasienergy spectra as a function of 
	$k_z$ in Figs.~\ref{3DSOTSC}~(g), (h), (i), and (j) when the system is in R1, R2, R3, and R4, respectively.
	\begin{figure}[]
		\centering
		\subfigure{\includegraphics[width=0.90\textwidth]{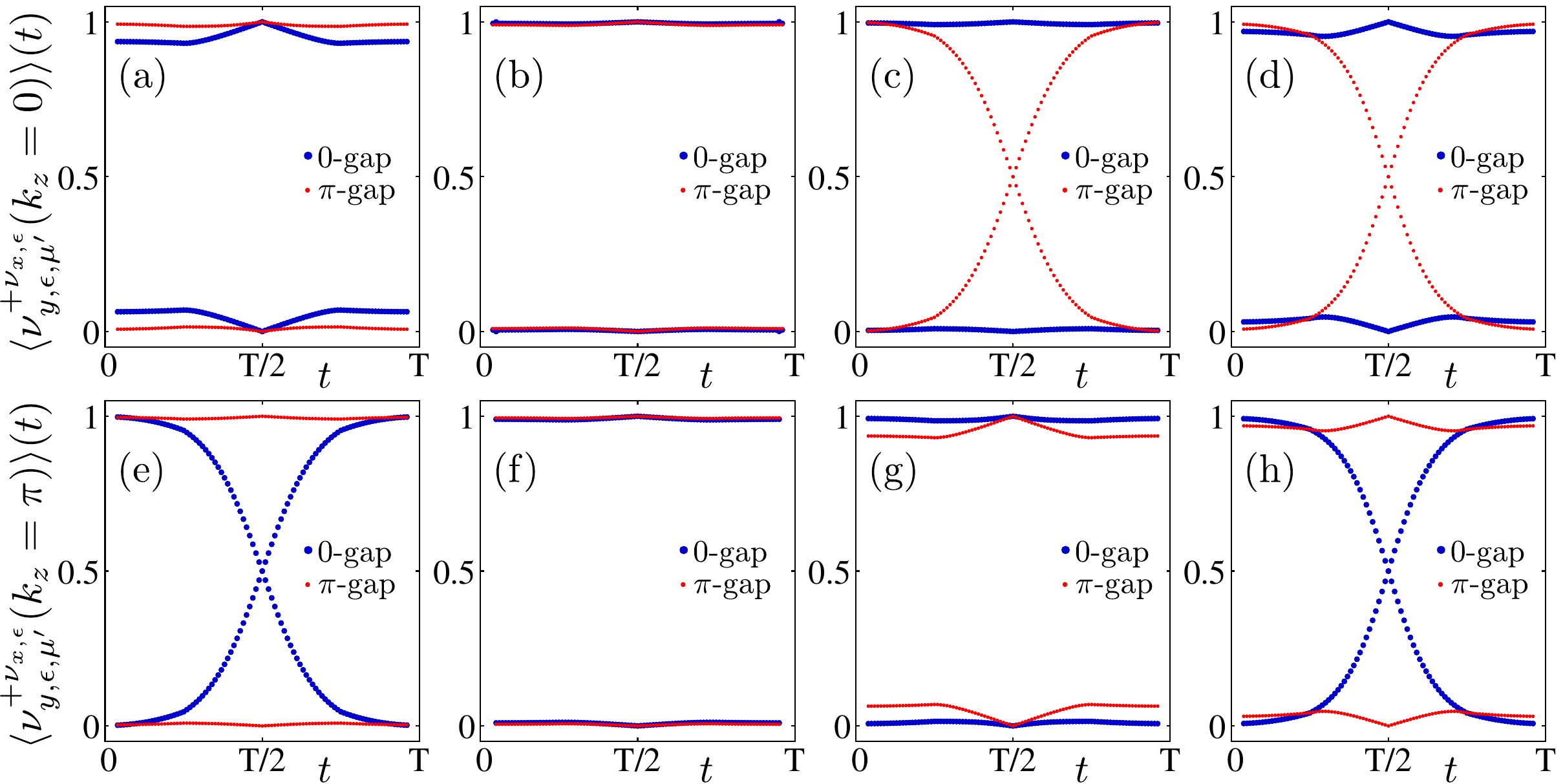}}
		\caption{We demonstrate the average quadrupolar motion $\langle \nu_{y,\epsilon , \mu'}^{+\nu_{x,\epsilon}} (k_z)\rangle (t)$ for 3D SOTSC at $k_z=0$ in panels (a), (b), (c), and (d), while the system 
			is in R1, R2, and R4, respectively. We repeat the same in panels (e)-(h) but considering $k_z=\pi$. Here, blue and red dots represent $\langle \nu_{y,\epsilon , \mu'}^{+\nu_{x,\epsilon}} (k_z)\rangle (t)$ arising from $0$ and $\pi$-gap, respectively as discussed in the text.
		}
		\label{3DSOTSCquad}
	\end{figure}
	
	Similar to the 2D FSOTSC, the 3D FSOTSC can be topologically characterized by the average quadrupolar motion except that we need to investigate the said quantity at a particular $k_z$. From Fig.~\ref{3DSOTSC}~(g)-(j), it is evident that the $0$-MHMs~(crosses through $E_m=0$) and the $\pi$-MHMs~(crosses through $E_m=\pi$) cross the corresponding quasi-energy at $k_z=\pi$ and $k_z=0$, respectively. Hence, $0$-MHMs~($\pi$-MHMs) can be topologically characterized by $\langle \nu_{y,\epsilon , \mu'}^{+\nu_{x,\epsilon}} (k_z=\pi)\rangle (t)$~($\langle \nu_{y,\epsilon , \mu'}^{+\nu_{x,\epsilon}} (k_z=0)\rangle (t)$). We depict $\langle \nu_{y,\epsilon , \mu'}^{+\nu_{x,\epsilon}} (k_z=0)\rangle (t)$~($\langle \nu_{y,\epsilon , \mu'}^{+\nu_{x,\epsilon}} (k_z=\pi)\rangle (t)$) in Figs.~\ref{3DSOTSCquad}~(a), (b), (c), and (d)~((e), (f), (g), and (h)) while the system is in R1, R2, R3, and R4, respectively. We obtain $0$-MHMs in R1 and R4~(see Figs.~\ref{3DSOTSC}~(g) and (j)). 
	Therefore, in these region $\langle \nu_{y,\epsilon , \mu'}^{+\nu_{x,\epsilon}} (k_z=\pi)\rangle (t)$ crosses $0.5~(\rm mod~1)$ as shown in Figs.~\ref{3DSOTSCquad}~(e) and (h)). Similarly, the 
	$\pi$-MHMs appear in R3 and R4~(see Figs.~\ref{3DSOTSC}~(i) and (j)) and in these region $\langle \nu_{y,\epsilon , \mu'}^{+\nu_{x,\epsilon}} (k_z=0)\rangle (t)$ crosses $0.5~(\rm mod~1)$~
	(see Figs.~\ref{3DSOTSCquad}~(c) and (d)). 
\end{onecolumngrid}

\end{document}